\documentclass[12pt]{article}
\usepackage{pdproc,epsfig}
\usepackage{graphicx}

\newcommand{\numu}{\mbox{$\nu_{\mu}$}}

\newcommand{\nue}{\mbox{$\nu_{e}$}}

\newcommand{\nueb}{\mbox{$\bar{\nu}_{e}$}}

\newcommand{\nutau}{\mbox{$\nu_\tau$}}
\newcommand{\nui}{\mbox{$\nu_1$}}
\newcommand{\nuii}{\mbox{$\nu_2$}}
\newcommand{\nuiii}{\mbox{$\nu_3$}}

\newcommand{\el}{\mbox{e$^{-}$}}

\newcommand{\Dm}{\mbox{$\Delta m^2$}}

\newcommand {\be}{\begin{equation}}
\newcommand {\ee}{\end{equation}}
\newcommand {\bea}{\begin{eqnarray}}
\newcommand {\eea}{\end{eqnarray}}
\newcommand{\bdec}{\mbox{$\beta$ decay}}
\newcommand{\nnbb}{\mbox{$0\nu \beta \beta$}}
\newcommand{\tnbb}{\mbox{$2\nu \beta \beta$}}
\newcommand{\belec}{\mbox{$\beta$ electron}}
\newcommand{\bspec}{\mbox{$\beta$ spectrum}}

\newcommand{\ezero}{\mbox{$E_0$}}
\newcommand{\mtwo}{\mbox{$m_\nu^2$}}
\newcommand{\dmtwo}{\mbox{$\Delta m_\nu^2$}}
\newcommand{\dee}{\mbox{$\Delta E/E$}}
\newcommand{\mnu}{\mbox{$m_\nu$}}
\newcommand{\mnue}{\mbox{$m(\nu_e)$}}
\newcommand{\ttwo}{\mbox{$\rm T_2$}}

\newcommand{\htwo}{\mbox{$\rm H_2$}}
\newcommand{\kr}{\mbox{$\rm ^{83m}Kr$}}
\newcommand{\bmax}{\mbox{$B_{\rm max}$}}
\newcommand{\bmin}{\mbox{$B_{\rm min}$}}
\newcommand{\ba}{\mbox{$B_{\rm A}$}}
\newcommand{\bso}{\mbox{$B_{\rm S}$}}
\newcommand{\bd}{\mbox{$B_{\rm D}$}}
\newcommand{\as}{\mbox{$A_{\rm S}$}}
\newcommand{\aseff}{\mbox{$A_{\rm S,eff}$}}
\newcommand{\aan}{\mbox{$A_{\rm A}$}}

\newcommand{\thmax}{\mbox{$\theta_{\rm max}$}}

\newcommand{\eg}{\mbox{\it e.g.}}
\newcommand{\ie}{\mbox{\it i.e.}}
\newcommand{\etal}{\mbox{\it et al.}}
\newcommand{\NIM}{\mbox{Nucl. Instr. and Meth.}}
\newcommand{\PRL}{\mbox{Phys. Rev. Lett.}}
\newcommand{\PL}{\mbox{Phys. Lett.}}
\newcommand{\PR}{\mbox{Phys. Rev.}}

\begin{document}
\pagestyle{empty}

\title{ {\large Letter of Intent} \\
    \vspace{1cm} {\large KATRIN: A next generation tritium beta decay
    experiment with sub-eV sensitivity for the electron neutrino mass}}
\author{A. Osipowicz$^a$,
H.~Bl\"umer$^{b,f}$,  G.~Drexlin$^b$, K.~Eitel$^b$, G.~Meisel$^b$,
P.~Plischke$^b$,  F.~Schwamm$^b$, M.~Steidl$^b$, H.~Gemmeke$^c$,
C.~Day$^d$, R.~Gehring$^d$, R.~Heller$^d$, K.-P.~J\"ungst$^d$,
P.~Komarek$^d$, W.~Lehmann$^d$, A.~Mack$^d$, H.~Neumann$^d$,
M.~Noe$^d$, T.~Schneider$^d$, L.~D\"orr$^e$, M.~Glugla$^e$,
R.~L\"asser$^e$, T.~Kepcija$^f$, J.~Wolf$^f$, J.~Bonn$^g$,
B.~Bornschein$^g$, L.~Bornschein$^g$, B.~Flatt$^g$, C.~Kraus$^g$,
B. M\"uller$^g$, E.W.~Otten$^g$, J.-P.~Schall$^g$, T.
Th\"ummler$^g$, C.~Weinheimer$^g$, V.~Aseev$^h$, A.~Belesev$^h$,
A.~Berlev$^h$, E.~Geraskin$^h$, A.~Golubev$^h$, O.~Kazachenko$^h$,
V.~Lobashev$^h$, N.~Titov$^h$, V.~Usanov$^h$, S.~Zadoroghny$^h$,
O.~Dragoun$^i$, A.~Koval\'{\i}k$^i$, M. Ry\v{s}av\'{y}$^i$,
A.~\v{S}palek$^i$,
    P.J.~Doe$^j$, S.R.~Elliott$^j$, R.G.H.~Robertson$^j$, J.F.~Wilkerson$^j$}

\address{ $^a$  University of Applied Sciences (FH) Fulda, Marquardtstr.~35,
         36039 Fulda, Germany}
\address{ $^b$ Forschungszentrum Karlsruhe, Institut f\"ur
Kernphysik, Postfach 3640, 76021 Karlsruhe, Germany}
\address{ $^c$ Forschungszentrum Karlsruhe, Institut f\"ur Prozessdatenverarbeitung und Elektronik, Postfach 3640, 76021 Karlsruhe, Germany}
\address{ $^d$ Forschungszentrum Karlsruhe, Institut f\"ur Technische Physik, Postfach 3640, 76021 Karlsruhe, Germany}
\address{ $^e$ Forschungszentrum Karlsruhe, Tritium-Labor Karlsruhe, Postfach 3640, 76021 Karlsruhe, Germany}
\address{ $^f$ Universit\"at Karlsruhe (TH),   Institut f\"ur Experimentelle Kernphysik,
    Gaedestr.~1, 76128 Karlsruhe, Germany}
\address{$^g$Johannes Gutenberg-Universit\"at Mainz, Institut f\"ur Physik,
         55099 Mainz, Germany}
\address{$^h$Academy of Sciences of Russia, Institute for Nuclear Research,
     60$\rm ^{th}$ October Anniversary Prospect 7a, 117312 Moscow, Russia}
\address{$^i$Academy of Sciences of the Czech Republic, Nuclear Physics Institute,
    CZ-250 68 \v{R}e\v{z} near Prague, Czech Republic}
\address{$^j$ Center for Experimental Nuclear Physics and Astrophysics, and
    Department of Physics, University of Washington, Seattle, WA 98195, USA}

\newpage
\abstract{
With the compelling evidence for massive neutrinos from recent $\nu$-oscillation
experiments, one of the most fundamental tasks of particle physics over the next
years will be the determination of the absolute mass scale of neutrinos.
The absolute value of $\nu$-masses  will have crucial implications for cosmology,
astrophysics and particle physics. We present the case for a next generation
tritium \bdec\ experiment  to perform a high precision direct measurement
of the absolute mass of the electron neutrino with sub-eV sensitivity.
We discuss the experimental requirements and technical challenges of the
proposed Karlsruhe Tritium Neutrino experiment (KATRIN) and
outline its physics potential.
}
\pagestyle{plain}
\newpage
\tableofcontents

\section{Introduction}

In this paper, we discuss the future of neutrino mass experiments
and present a plan for a large tritium \bdec\ experiment with
sub-eV sensitivity to the mass of the electron neutrino. The
structure of the article is as follows: In the introduction, we
report briefly the evidence for neutrino masses, the implications
of non-zero neutrino masses for particle physics and cosmology,
and discuss several experimental approaches to determine neutrino
masses. The current tritium \bdec\ experiments in Mainz/Germany
and Troitsk/Russia are described in section 2. In section 3, we
present an outline of the proposed future tritium \bdec\
experiment KATRIN.

\subsection{Evidences for massive neutrinos}

In modern particle physics, one of the most intriguing and most
challenging tasks is to discover the rest mass of neutrinos,
bearing fundamental implications to particle physics, astrophysics
and cosmology. Until recently, the Standard Model (SM) of particle
physics assumed neutrinos to be massless. However, actual
investigations of neutrinos from the sun and of neutrinos created
in the atmosphere by cosmic rays, in particular the recent results
of the Super-Kamiokande and SNO experiments
\cite{sk-atm,sk-solar,snocc}, have given strong evidence for
massive neutrinos indicated by neutrino oscillations. Neutrino
oscillations imply that a neutrino from one specific weak
interaction flavor, \eg\ a muon neutrino \numu , transforms into
another weak flavor eigenstate, \ie\ an electron neutrino \nue\ or
a tau neutrino \nutau , while travelling from the source to the
detector. The existence of neutrino oscillations requires a
non-trivial mixing between the weak interaction eigenstates (\nue
, \numu , \nutau ) and the corresponding neutrino mass states
(\nui , \nuii , \nuiii ) and, moreover, that the mass eigenvalues
($m_1, m_2, m_3$) differ from each other. Consequently, the
experimental evidence for neutrino oscillation proves that
neutrinos have non-zero masses.

\subsection{Implications of neutrino masses}
The existence of neutrino oscillations and therefore of neutrino
mixing and masses has far-reaching implications to numerous fields
of particle physics, astrophysics and cosmology:
\begin{itemize}
\item {\it Particle Physics:}

The SM of particle physics, which very precisely describes the present
experimental data up to the electroweak scale, offers no explanation
for the observed pattern of the fermion masses or the
mixing among the fermion generations. In particular, it offers no
explanation for neutrino masses and neutrino mixing. Accordingly,
the recent experimental evidence for neutrino masses and
mixing is the first indication for physics beyond the Standard Model.

There are many theories beyond the Standard Model, which explore the
origins of neutrino masses and mixings. In these theories, which often work
within the framework of Supersymmetry, neutrinos naturally acquire mass.
A large group of models makes use of the so-called see-saw effect
to generate neutrino masses \cite{seesaw}. Other classes of theories are
based on completely different possible origins of neutrino masses, such
as radiative corrections arising from an extended Higgs sector \cite{radiative}.
As neutrino masses are much smaller than the masses of the other
fermions, the knowledge of the absolute values of neutrino masses is
crucial for our understanding of the fermion
masses in general. Recently it has been pointed out  that the
{\it absolute mass scale of neutrinos} may
be even more significant and straightforward for the fundamental
theory of fermion masses than the determination of the neutrino
mixing angles and CP-violating phases \cite{smirnov}. It will most
probably be the absolute mass scale of neutrinos which will determine
the scale of new physics.

All these theories extended beyond the SM can be grouped into two
different classes, leading either to a hierarchical pattern for
the neutrino mass eigenvalues $m_i$ (following the pattern of the
quark and charged lepton masses)
\begin{equation}
m_1 \ll m_2 \ll m_3
\end{equation}
or resulting in a nearly degenerate pattern of neutrino masses
\begin{equation}
m_1 \approx m_2 \approx m_3
\end{equation}

As neutrino oscillation experiments are only sensitive to the
differences of the squared masses $\Delta m^2$ (\eg\ $\Delta
m^2_{12}=|m_1^2-m_2^2|$), they cannot measure absolute values of
$\nu$ masses. While they do not distinguish between the two
classes of models, oscillation experiments allow to set a {\it
lower bound} on the absolute $\nu$-mass, as at least one of the
neutrino mass eigenvalues should satisfy the inequality\,:
\begin{equation}
m_i \geq \sqrt{|\Delta m^2|}
\end{equation}
Analysis of the actual results of Super-Kamiokande \cite{sk-atm} in terms
of oscillations of atmospheric neutrinos thus gives a lower bound on $m_3$:
\begin{equation}
m_3 \geq \sqrt{\Delta m^2_{atm}} \sim (0.04 -  0.07) ~ {\rm eV}
\end{equation}
However, the fundamental mass scale of neutrinos can be located
orders of magnitude above this lower bound  (\eg\ at around
1\,eV), as suggested by mass models with degeneracy
\cite{degenerate}. Discrimination between hierarchical and
degenerate mass scenarios thus requires a sensitivity on the
absolute $\nu$-mass scale in the sub-eV range.

Finally, theoretical models come to different conclusions of
whether neutrino masses are of the Dirac- or of the Majorana type.
A massive neutrino which is identical to its own antiparticle is
called a Majorana particle, while for Dirac-type neutrinos the
lepton number distinguishes neutrinos from antineutrinos. This
requires the development of experimental techniques for
$\nu$-masses in the sub-eV range, which do not depend on
assumptions about the Dirac or Majorana character of the neutrino
mass.

\item {\it Cosmology and Astrophysics:}

In astrophysics and cosmology, neutrino masses and mixings play an
important role in numerous scenarios, ranging from the formation
of light nuclei during the Big Bang nucleosynthesis, the formation
and evolution of large scale structures in the universe, up to
stellar evolution and the very end of a heavy star, a supernova
explosion \cite{Fuller}. Of special interest are the relic
neutrinos left over from the Big Bang. The number of these
neutrinos in the universe is huge, equivalent to the photons of
the Cosmic Microwave Background Radiation (CMBR). The ratio of
relic neutrinos to baryons is about 10$^9$:1, therefore even small
neutrino masses are of great importance.

Thus neutrinos could contribute in a significant way to the large
amount of dark matter in the universe. In this context it is
important to have in mind that neutrinos act as socalled
relativistic or Hot Dark Matter (HDM), whereas other dark matter
candidates (supersymmetric particles) represent non-relativistic
or Cold Dark Matter (CDM). Cosmological models of structure
formation strongly depend on the relative amounts of cold and hot
dark matter in the universe, hence a determination of the neutrino
HDM contribution to the total dark matter content of the universe
would be important for our understanding of structure formation
\cite{kamion}.

\begin{figure}
 \epsfxsize=10.6cm
 \centerline{\epsffile{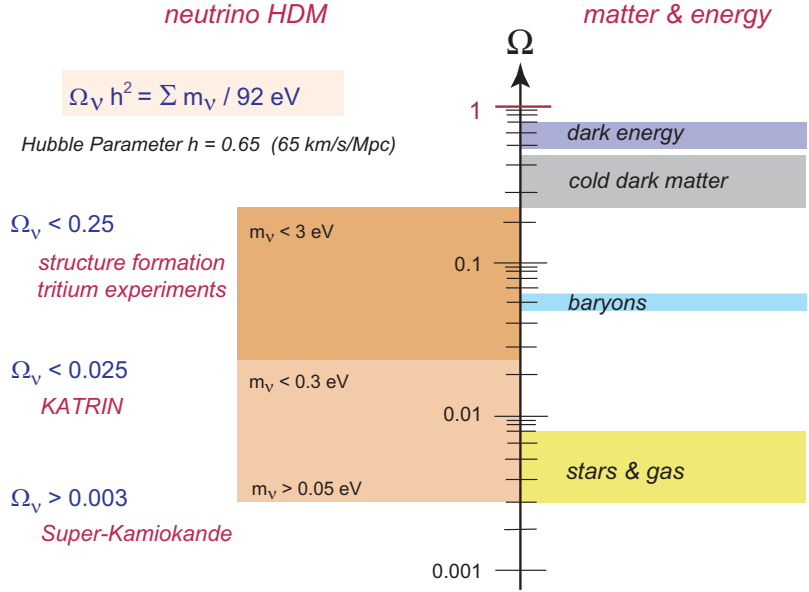}}
 \caption{The contribution $\Omega_\nu$ from neutrino HDM to the total matter
 energy density $\Omega$ of the universe spans two orders of magnitude. The
 lower bound on $\Omega_\nu$ comes from the analysis of oscillations of
 atmospheric $\nu$'s. The upper bound  stems from current tritium \bdec\
 experiments and studies of structure formation. The main motivation of
 the proposed tritium \bdec\ experiment KATRIN (see section~\ref{chap3})
 will be to investigate the $\Omega_\nu$ interval from 0.025-0.25,
 where the relic neutrinos from the Big Bang would play a
 significant role as $\nu$ HDM in the evolution of large scale structures.
  }
 \label{omega_nu}
 \end{figure}

Fig. 1 shows the different contributions to the total
matter-energy density $\Omega$ of the universe, arising from
luminous matter, baryons, CDM and the so-called Dark Energy (which
is equivalent to the cosmological constant $\Lambda$). While the
individual contributions of these components are more or less well
determined, the contribution $\Omega_\nu$ of neutrino HDM can vary
in the interval $ 0.003 < \Omega_\nu < 0.25$.  The lower bound on
$\Omega_\nu$ arises from the results of Super-Kamiokande on the
oscillations of atmospheric neutrinos \cite{sk-atm}. The upper
bound comes from current tritium \bdec\ experiments \cite{pdg00}
and, independently, from recent studies of the evolution of large
scale structures in the universe making use of the solar and
atmospheric oscillation results \cite{tegm}. The parameter range
of $\Omega_\nu$ for neutrino HDM, which is presently allowed by
experiment, thus spans two orders of magnitude. Clearly, the
present situation with regard to $\Omega_\nu$ is not satisfactory.
A determination of  $\Omega_\nu$ or a significant constraint on
the allowed parameter range of  $\Omega_\nu$ would lead to a much
better understanding of the role of neutrino HDM in the evolution
of large scale structures.

One means of identifying $\nu$ HDM and constraining $\Omega_\nu$
are precise measurements of the temperature fluctuations of the
CMBR with balloon or satellite experiments. However, the
interpretation of these data crucially depends on model
assumptions and the precise knowledge of other cosmological
parameters. If, on the other hand, the absolute mass scale of
neutrinos could be determined with sub-eV precision by a
laboratory experiment, the corresponding $\Omega_\nu$ would be a
\textit{fixed} parameter in the analysis of CMBR experiments. This
would be especially important for the analysis of high precision
CMBR experiments like the recently started MAP \cite{MAP}
satellite and the future PLANCK mission \cite{PLANCK}.

The investigation of the still open role of neutrino HDM in the
evolution of large scale structure is  one of the main motivations
for the proposed next-generation tritium \bdec\ experiment KATRIN,
which is designed to measure the absolute mass of the electron
neutrino with sub-eV sensitivity. Correspondingly, KATRIN would be
sensitive to a neutrino HDM contribution down to a value of
$\Omega_\nu$= 0.025, thus significantly constraining the role of
neutrino  HDM in structure formation.

\end{itemize}

As discussed above, the absolute mass scale of neutrinos plays an important role
in astrophysics, cosmology and particle physics. While a series of future oscillation
experiments using solar \cite{solar}, reactor \cite{KamLand} and accelerator neutrinos
\cite{K2K,Minos,BooNE} will improve our understanding of
neutrino mixing, the problem of setting the absolute scale of neutrino masses will remain
and become a key issue in particle physics.

This absolute mass scale can be inferred by two different experimental approaches:
the search for the so-called neutrinoless double \bdec\ and the direct
kinematic neutrino mass experiments. Both types of experiments are measuring
different neutrino mass parameters and hence are complementary to each other.

\subsection{The search for the neutrinoless double \bdec}
The search for the neutrinoless double \bdec\ (\nnbb)
is a very sensitive means of searching for neutrino masses. The physical process is a
twofold \bdec\ in one nucleus at the same time. The normal double \bdec\ with the
emission of two electron neutrinos (or electron antineutrinos) \tnbb\ was observed
already 14 years ago, but yields no information on the neutrino mass.
In the case of the neutrinoless double \bdec , the neutrino emitted at one \bdec\
vertex has to be absorbed at the second decay vertex as an antineutrino. This
process will only take place on condition that the neutrino is massive and identical
to its own antiparticle, \ie\ it has to be a Majorana particle. Correspondingly,
this process violates lepton  number conservation by two units.

Up to now, neutrinoless double \bdec\ has not been observed. For
the isotope $^{76}{\rm Ge}$, the current best half life limit from
the Heidelberg-Moscow experiment results in an upper limit on the
effective Majorana mass of the electron neutrino of $m_{ee} <
0.34$\,eV at 90\% confidence \cite{HdM}. There are several
projects aiming at increasing the sensitivity of \nnbb\ searches
into the range of below 0.1\,eV \cite{fiorini}. It is evident that
these limits apply only to Majorana neutrino masses and not to
Dirac type masses (see sect. 1.5 for the meaning of $m_{ee}$ and
its relation to the true neutrino mass scale).

\subsection{Direct investigations of the neutrino masses}
In contrast to double \bdec\ experiments,
direct investigations of the neutrino mass do not rely on further assumptions on
the neutrino mass type (Majorana or Dirac). Direct or kinematic experiments can be
classified into two categories both making use of the
relativistic energy momentum relation $E^2 = p^2 c^2 + m^2 c^4$ as well as of
energy and momentum conservation.

\subsubsection{Time-of-flight method}
Using a time-of-flight method as a means of measurement of the
neutrino masses requires very large distances between source and
detector and therefore very intense neutrino sources which only
astrophysical, cataclysmic events can provide. The observation of
some 20 neutrinos from the Supernova 1987A yielded an upper limit
of $m(\nue)<23$\,eV \cite{Loredo} by measuring the correlation
between the energy and the arrival time of the supernova
neutrinos. Though a considerable improvement of the above number
can be expected from the measurement of neutrinos from a future,
nearby galactic supernova by large underground neutrino
experiments (\eg\ Super-Kamiokande, SNO), the expected sensitivity
will not reach a value below 1\,eV \cite{Beacom}. Moreover, the
estimated rate of supernova type II explosions is only in the
range of one event per several tens of years in our galaxy.

\subsubsection{Kinematics of weak decays}
The investigation of the kinematics of weak decays is based on the measurement of
the charged decay products of weak decays.
For the masses of \numu\ and \nutau\ the measurement of pion decays
into muons and \numu\ at PSI and the investigation of $\tau$-decays into 5 pions
and \nutau\ at LEP have yielded the upper limits:
\begin{eqnarray*}
  m(\numu) &<& 190~\mbox{ keV~~~~~~ at~ 90\, \%~ confidence ~\cite{mmu}} \\
  m(\nutau) &<& 18.2~\mbox{ MeV~~~~ at~ 95\, \%~ confidence ~\cite{mtau}}
\end{eqnarray*}
Both limits are much larger than the interesting range for
cosmology and $\nu$ HDM (see fig.~\ref{omega_nu}). However,
experiments investigating the mass of the electron neutrino \nue\
by analyzing \bdec s with emission of electrons are providing a
sensitivity in the interesting eV-range (see
section~\ref{tritdec}).

\subsubsection{Tritium \bdec}\label{sstrit}

The most sensitive direct searches for the electron neutrino mass up to now
are based on the investigation of the electron spectrum of tritium \bdec\
\begin{displaymath}
  ^3{\rm H}\,\rightarrow \, ^3{\rm He}^+\,+\,\el \,+\,\nueb \quad {\rm .}
\end{displaymath}

The electron energy spectrum of tritium \bdec\ for a
neutrino with mass $m_\nu$ is given by
\begin{equation}
{dN \over dE} = C \times F(Z,E) p E(E_0-E) [(E_0-E)^2-m_\nu^2]^{1
\over 2} \Theta (E_0-E-m_\nu), \label{mother}
\end{equation}
where $E$ denotes the electron energy, $p$ is the electron
momentum,  $E_0$ corresponds to the total decay energy, $F(Z,E)$
is the Fermi function, taking into account the Coulomb interaction
of the outgoing electron in the final state, the stepfunction
$\Theta (E_0-E-m_\nu)$ ensures energy conservation, and $C$ is
given by \be C=G_F^2 {m_e^5  \over 2 \pi^3} \cos^2 \theta_C |M|^2
~. \label{re} \ee Here $G_F$ is the Fermi constant, $\theta_C$ is
the Cabibbo angle, $m_e$ the mass of the electron and $M$ is the
nuclear matrix element. As both  $M$ and $F(Z,E)$ are independent
of $m_\nu$, the dependence of the spectral shape on $m_\nu$ is
given by the phase space factor only. In addition, the bound on
the neutrino mass from tritium \bdec\ is independent of whether
the electron neutrino is a Majorana or a Dirac particle.

\begin{figure}
  \epsfxsize=15cm
  \centerline{\epsffile{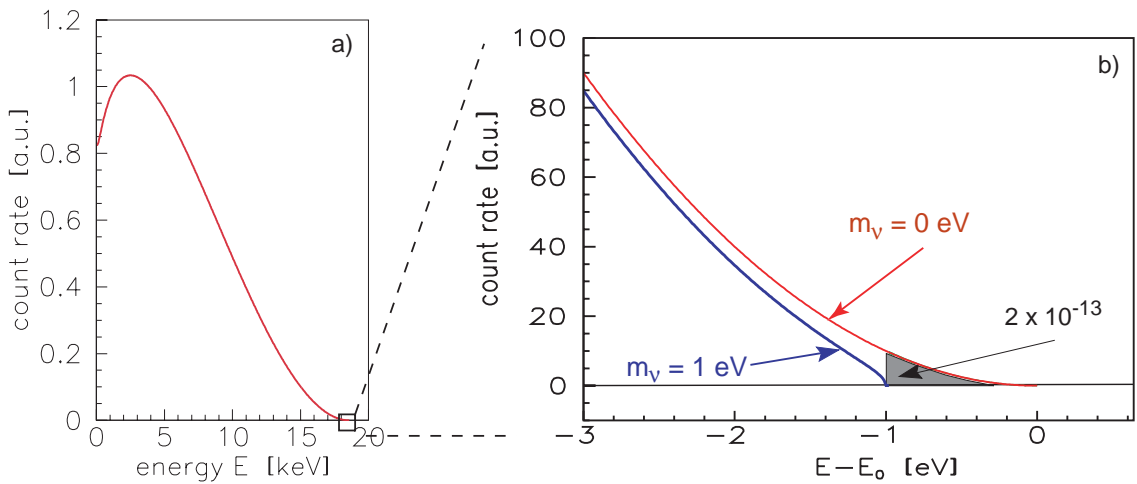}}
  \caption{ The electron energy spectrum of tritium \bdec : (a) complete and (b) narrow
   region around endpoint \ezero . The \bspec\ is shown for neutrino
   masses of 0 and 1~eV.
  }
  \label{fig_betaspec}
  \end{figure}

The signature of an electron neutrino with a mass of
m(\nue\,)=1\,eV is shown in fig. 2 in comparison with the
undistorted $\beta$ spectrum of a massless \nue\,. The spectral
distortion is statistically significant only in a region close to
the $\beta$ endpoint. This is due to the rapidly rising count rate
below the endpoint ${\rm dN/dE} \propto (E_0 - E)^2$. Therefore,
only a very narrow region close to the endpoint $E_0$ is analyzed.
As the fraction of \bdec s in this region is proportional to a
factor $(1 / E_0)^3$, the very low tritium endpoint energy of 18.6
keV maximizes the fraction of \bdec s in this region (in fact,
tritium has the second lowest endpoint of all $\beta$ unstable
isotopes). Nevertheless, the requirements of tritium \bdec\
experiments with regard to source strength are demanding. As an
example, the fraction of \bdec s falling into the last 1\,eV below
the endpoint $E_0$ is only 2 $\times$ 10$^{-13}$ (see
fig.~\ref{fig_betaspec}), hence tritium \bdec\ experiments with
high neutrino mass sensitivity require a huge luminosity combined
with very high energy resolution.

Apart from offering a low endpoint energy $E_0$ and a moderate
half life of 12.3\,y, tritium has further advantages as $\beta$
emitter in $\nu$ mass investigations:
\begin{enumerate}
\item the hydrogen isotope tritium and its daughter, the $^3$He$^+$ ion,
      have a simple electron shell configuration.  Atomic corrections
       for the \bdec ing atom -or molecule- and corrections due to the interaction
      of the outgoing \belec\ with the tritium source can be calculated in a simple and
      straightforward manner
\item The tritium \bdec\ is a super-allowed nuclear transition. Therefore, no corrections
      from the nuclear transition matrix elements $M$ have to be taken into account.
\end{enumerate}
The combination of all these features makes tritium an almost ideal $\beta$ emitter
for neutrino mass investigations.

\subsubsection{Other approaches to \bdec}
A different approach to directly measure the electron neutrino
mass is the use of cryogenic bolometers. In this case, the $\beta$
source can be identical to the \belec\ spectrometer. This new
technique has been applied to the isotope $^{187}{\rm Re}$, which
has a 7 times lower endpoint energy than tritium\cite{Milano}. The
experiments are still in the early stage of development. Current
microcalorimeters reach an energy resolution of $\Delta E \sim
5$\,eV for short-term measurements and yield an upper limit of
$m(\nue)<26$\,eV\ \cite{gatti}. To further improve the statistical
accuracy, the principle of integration of active source and
detector requires the operation of large arrays of
microcalorimeters. The expected sensitivity on the neutrino mass
in the nearer future will be below $\sim 10$\,eV \cite{gatti}.

\subsection{Neutrino mixing and mass scale}

The effects of neutrino mixing on tritium \bdec\ and \nnbb\ experiments
can be significant. Below we discuss the implications of $\nu$-mixing in the
determination of the absolute mass scale of neutrinos for both experimental
approaches, following largely the discussions in \cite{smirnov,petcov}.

\subsubsection{Mixing and \nnbb}

Considering neutrino mixing
for \nnbb\ decay, the {\it effective} Majorana mass $m_{ee}$
is  a coherent sum of all neutrino mass states $\nu_i$ contributing
to the electron neutrino \nue . The parameter $m_{ee}$ is a combination of
mass eigenvalues $m_i$, Majorana phases and mixing parameters given by
\begin{equation}
\label{eff_mass_bb}
  m_{ee} = \left| \sum_{i=1}^{3} U^2_{ei} \cdot m_i \right|
\end{equation}
Since the Majorana CP-phases are unknown, strong cancellations in the sum
over all neutrino states $\nu_i$ can occur.

\subsubsection{Mixing and tritium \bdec}

In the case of tritium \bdec , the presence of mixing modifies eq.
(\ref{mother}) to: \be {dN \over dE}=C \times F(Z,E) p E (E_0-E)
\sum_i |U_{ei}|^2 [(E_0-E)^2-{m_i}^2]^{1 \over 2} \Theta
(E_0-E-m_i), \label{improve} \ee The step function,
$\Theta(E_0-E-m_i)$, ensures that a neutrino state $\nu_i$ is only
produced if the energy available is larger than its mass. In
general, the effects of mixing will lead to the following spectral
modifications :
\begin{enumerate}
\item the $\beta$ spectrum ends at $E_{0'} = E_0 - m_1$, where $m_1$ is the lightest
mass in the neutrino mass spectrum (i.e. the electron spectrum bends at
$E \stackrel {<} {\sim} E_{0'}$)
\item the appearance of 'kinks' at the electron energy $E_e^i \approx E_0 - m_i$,
with the size of the kinks being determined by the mixing elements $|U_{ei}|^2$.
\end{enumerate}

For general mixing schemes with 3~neutrinos (3$\nu$) or 3~active
neutrinos and 1~'sterile' (4$\nu$), the spectral shape of tritium
\bdec\ can be rather complex, requiring the introduction of at
least five independent parameters (two mixing parameters and three
masses for 3$\nu$-mixing). In \cite{smirnov,petcov} all possible
3$\nu$ or 4$\nu$ mixing schemes (with normal or inverted mass
hierarchy) have been discussed extensively. In the following, we
restrict our discussion to the schemes explaining the solar and
atmospheric neutrino oscillation data. This omits the 4$\nu$
schemes incorporating the LSND oscillation results \cite{LSND},
which, however, have not been confirmed by other experiments such
as KARMEN \cite{KARMEN}.

If the pattern of neutrino masses is hierarchical ($m_1 \ll m_2 \ll m_3 $),
the largest mass, $m_3 \simeq \sqrt{\Delta m_{atm}^2}=(4-7) \times 10^{-2}$ eV,
would be too small to be observed in tritium $\beta$ decay. For models of
quasi-degenerate neutrino masses with an absolute mass scale in the range of
sensitivity of future tritium $\beta$ experiments, the effects of non-zero $\nu$-masses
and mixings reduce to a single parameter, $m^2(\nue)$. With $m_1 \approx m_2 \approx m_3$,
the only distortion of the spectrum to be seen is a bending at the
electron energy $E_0 - m(\nue)$, equivalent to the case of a \nue\ with definite mass and
no mixing. Therefore the analysis of the $\beta$ spectrum can be parametrized by
\begin{equation}
  \label{eff_mass_tri}
  m^2(\nue) = \sum_{i=1}^{3} |U_{ei}|^2 \cdot m^2_i \quad {\rm .}
\end{equation}
In contrast to eq. (\ref{eff_mass_bb}), here the absolute values
of the squared mixing matrix elements are involved. Therefore the
sum contains only non-negative elements, no cancellations can
happen. Hence, the neutrino mass $m(\nue)$ extracted from the
experiment fixes the absolute mass scale ($m_1 \approx m_2 \approx
m_3$), taking into account the small values of \Dm\ from
oscillation experiments.

\subsubsection{Absolute neutrino mass scale}

While tritium \bdec\ and \nnbb\ are largely complementary to each
other, it is nevertheless interesting to compare the two
parameters $m(\nue)$ and $m_{ee}$ with each other (assuming
neutrinos are Majorana particles) and to investigate their
relation to the fundamental neutrino mass scale in the presence of
$\nu$-mixing. For a 3$\nu$ mixing the following bounds on the
\bdec\ mass $m(\nue)$ can be derived\,:  \be m_{ee} < m(\nue) <
{m_{ee} \over \left| |\cos 2 \theta_\odot|(1-|U_{e3}|^2) -
|U_{e3}|^2 \right| } , \label{bound} \ee The \nue\ mixing
parameters in (\ref{bound}) can be deduced from the results of
oscillation experiments using solar neutrinos (solar mixing angle
$\theta_\odot$) and reactor antineutrinos ($|U_{e3}|^2$).

 \begin{figure}
  \epsfxsize=12.cm
  \centerline{\epsffile{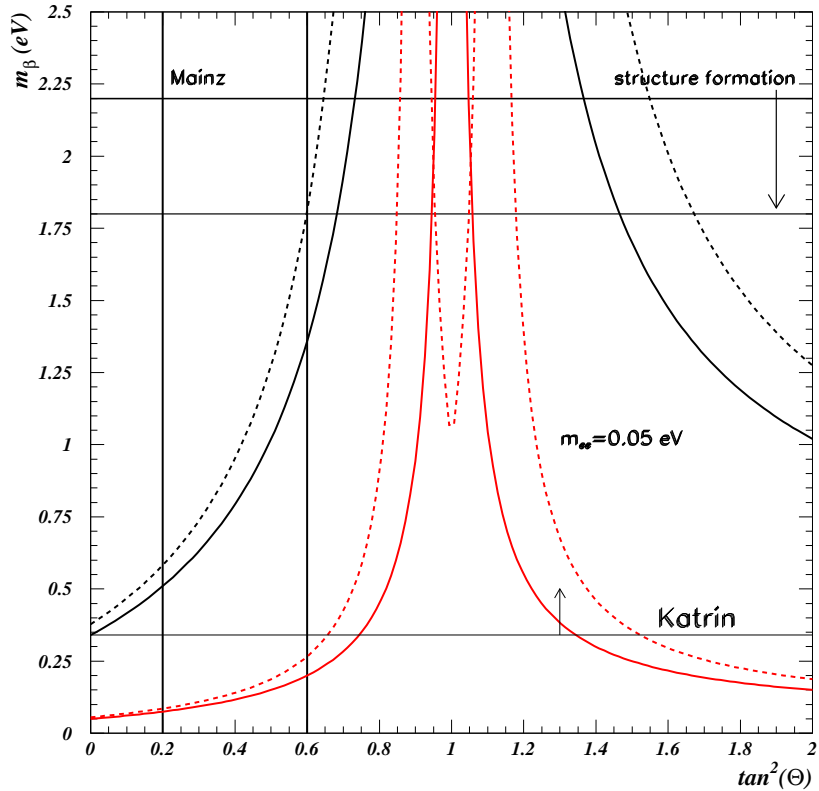}}
  \caption{Bounds on the effective \bdec\ mass $m(\nue)$ ($=m_{\beta}$ in
  ref.~\protect\cite{smirnov}) as functions of the solar mixing
  angle $\theta_{\odot}$ (for 3$\nu$ mixing models with strong mass
  degeneracy). The
  horizontal lines correspond to the current and the proposed
  future tritium \bdec\ experiments (Mainz and KATRIN) as well as to structure formation.
  For results from \nnbb\ experiments, which strongly depend on $\theta_{\odot}$,
  the upper solid (dashed) line corresponds to the
  present bound $m_{ee}\leq 0.34$\,eV and $|U_{e3}|^2=0$
  ($|U_{e3}|^2$=0.05), the lower solid (dashed) line corresponds
  to an envisaged future limit of
  $m_{ee} \leq  0.05$\,eV  and $|U_{e3}|^2=0$ ($|U_{e3}|^2$=0.05).
  The vertical lines mark the current 90\,\% C.L. borders of the large mixing angle (LMA)
  solution region for solar $\nu$'s.
  (fig. taken from ref.~\protect\cite{smirnov}) }
  \label{mass_scales}
  \end{figure}

Fig.~\ref{mass_scales} shows the bounds on $m(\nue)$ from current
and future tritium \bdec\ and \nnbb\ experiments as a function of
the solar mixing angle  $\theta_\odot$ for two different values of
$|U_{e3}|^2$.  It is interesting to note that for large values of
$\theta_{\odot}$ a positive signal from \nnbb\ experiments only
yields a {\it lower limit} on the fundamental neutrino mass scale.
The relationship between $m_{ee}$ and the 'true' mass scale
$m(\nue)$ depends crucially on the solar mixing parameter
$\theta_\odot$. The weakest \nnbb\ bounds on $m(\nue)$ appear at
maximum mixing of solar neutrinos $\tan^2 \theta_{\odot} = 1$ (for
$|U_{e3}|^2$=0). The actual allowed parameter spaces for
$\theta_\odot$ from solar neutrino experiments (including the
recent SNO results) have been evaluated in \cite{lisi}. While
small solar mixing angles $\tan^2 \theta_{\odot} < 0.01$
(so-called SMA solutions) are disfavored by current results,
solutions with large mixing angle $\tan^2 \theta_{\odot} > 0.1$
(either the socalled LMA or LOW solutions) are strongly
 favored.

This implies that the absolute $\nu$-mass scale cannot easily be restored from $m_{ee}$ due
to possible cancellations depending on the unknown CP-violating phases. Moreover, there are
uncertainties in the calculations of the nuclear matrix elements of \nnbb\ transitions, which may
lead to systematic errors in the relation between $m_{ee}$ and the true mass scale. Finally,
many other exchange modes apart from light Majorana particles can contribute to the
\nnbb\ transition rate \cite{klapdor}, thereby adding further uncertainties.
Thus the development of direct methods of determination of the neutrino mass is essential
for a complete reconstruction of the neutrino mass spectrum.

\section{Tritium \bdec\ experiments \label{tritdec}}

The almost ideal features of tritium as a $\beta$ emitter have
been the reason for a long series of tritium \bdec\ experiments.
\begin{figure}
\epsfxsize=10.5cm \centerline{\epsffile{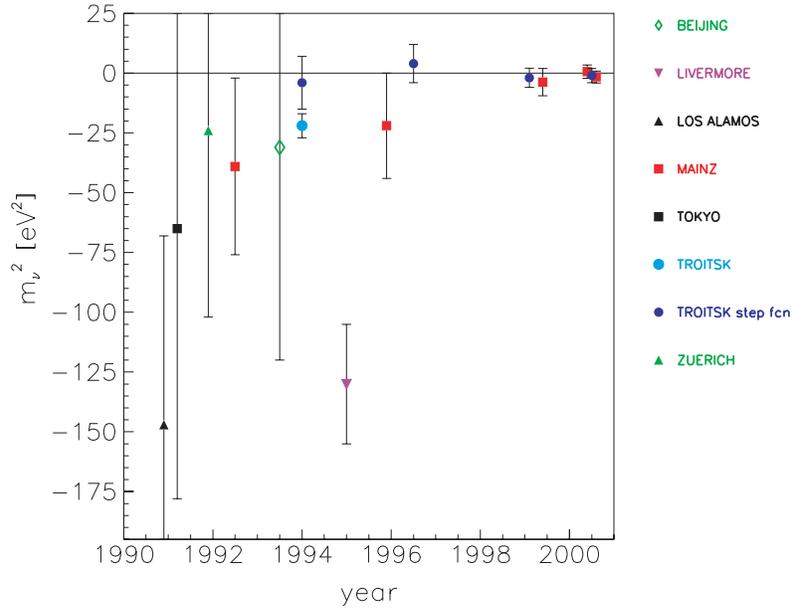}}
  \caption{Results of tritium \bdec\ experiments on the observable
  \mtwo\
   over the last decade. The experiments at Los Alamos, Z\"urich,
   Tokyo, Beijing and Livermore \protect\cite{LANL,Zuerich,Tokyo,Bejing,LLNL}
   used magnetic spectrometers, the
   experiments at Troitsk and Mainz \protect\cite{Lobashev99,Weinheimer99} are using
   electrostatic spectrometers of the MAC-E-Filter type (see text).}
  \label{fig_tritiumexp}
\end{figure}
Figure \ref{fig_tritiumexp} shows the evolution on the observable
\mtwo\ of the various tritium \bdec\ experiments over the last
decade. It is remarkable that the error bars on \mtwo\ have
decreased by nearly two orders of magnitude. Equally important is
the fact that the problem of negative values for \mtwo\ of the
early nineties has disappeared due to better understanding of
systematics and improvements in the experimental setups.

\subsection{MAC-E-Filter}
The high sensitivity of the Troitsk and the Mainz neutrino mass
experiments is due to a new type of spectrometers, so-called
MAC-E-Filters (\underline{M}agnetic \underline{A}diabatic
\underline{C}ollimation combined with an
  \underline{E}lectrostatic Filter). This new type was first proposed in
  \cite{beamson}. Later this method was reinvented specifically
  for the search for the electron neutrino mass at Troitsk and Mainz,
  independently. It combines high luminosity
and low background with a high energy resolution, both essential
to measure the neutrino mass from the endpoint region of a \bdec\
spectrum.
  \begin{figure}
  \epsfxsize=10cm
  \centerline{\epsffile{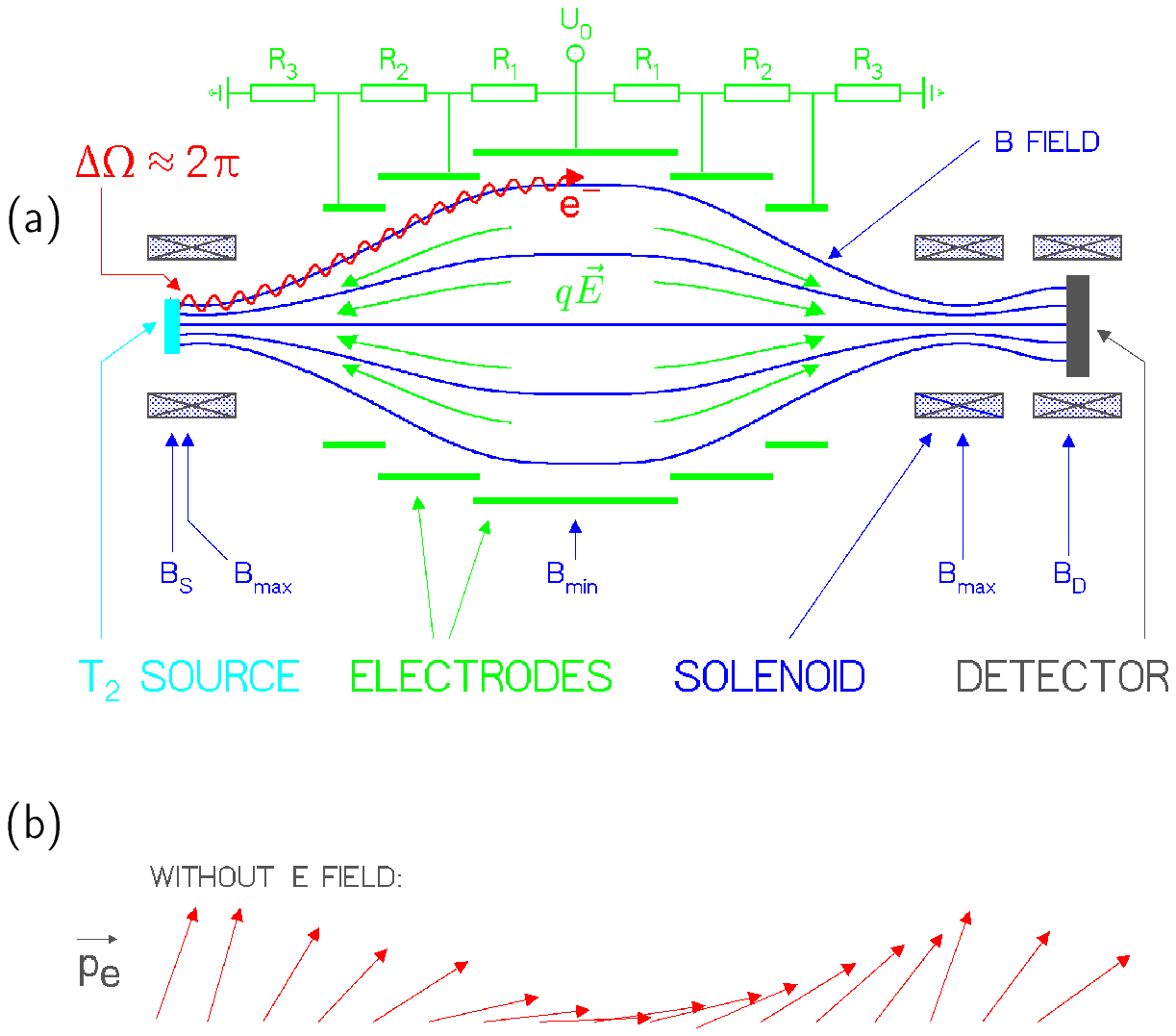}}
  \caption{Principle of the MAC-E-Filter. (a) Experimental setup, (b) Momentum
   transformation due to adiabatic invariance of magnetic orbit momentum $\mu$ in the
   inhomogeneous magnetic field.}
  \label{fig_mace}
  \end{figure}

  The main features of the MAC-E-Filter are illustrated in fig. \ref{fig_mace}(a).
  Two superconducting solenoids are producing a magnetic guiding field.
  The \belec s, which are starting from the tritium source
  in the left solenoid into the forward hemisphere, are
  guided magnetically on a cyclotron motion around the magnetic field lines
  into the spectrometer, thus resulting in an accepted solid
  angle of up to $2 \pi$.
  On their way into the center of the spectrometer the magnetic
  field $B$ drops by many orders of magnitude. Therefore,
  the magnetic gradient force transforms most of the cyclotron energy $E_\perp$
  into longitudinal motion. This is illustrated in fig. \ref{fig_mace}(b)
  by a momentum vector. Due to the slowly varying magnetic field the
  momentum transforms adiabatically, therefore the magnetic moment $\mu$
  keeps constant
  (equation given in non-relativistic approximation):
  \begin{equation}
    \mu = \frac{E_\perp}{B} = const.
  \end{equation}
  This transformation can be summarized as follows:
  The \belec s, isotropically emitted at the source, are
  transformed into a broad beam of electrons flying almost parallel to the
  magnetic field lines.

  This parallel beam of electrons is running against an
  electrostatic potential formed by a system of cylindrical electrodes.
  All electrons with enough energy to pass the electrostatic barrier
  are reaccelerated and collimated onto a detector, all others are reflected.
  Therefore the spectrometer acts as an integrating high-energy pass filter.
  The relative sharpness \dee\ of this filter is given by the ratio of the minimum
  magnetic field $B_{A}$ in the center plane and the maximum magnetic field $B_{max}$ between
  \belec\ source and spectrometer :
  \begin{equation}
    \frac{\Delta E}{E} = \frac{B_A}{B_{max}} \quad {\rm .}
    \label{eq_resolution}
  \end{equation}
  Varying the electrostatic retarding potential allows to measure the \bspec\  in an
 integrating mode.

In order to suppress electrons which have a very long path within
the tritium source and therefore possess a high scattering
probability, the electron source is placed in a magnetic field
$B_S$ (see fig.~\ref{fig_mace}), which is lower than the maximum
magnetic field \bmax\,. This restricts
 the maximum accepted starting angle of the electrons
\thmax\ by the magnetic mirror effect to\,:
\begin{equation}
  \sin \thmax = \sqrt{\frac{\bso }{\bmax }}
  \label{eq_thetamax}
\end{equation}

\subsection{The Mainz and the Troitsk experiments}
  The experiments at Troitsk \cite{Lobashev99} and Mainz \cite{Weinheimer99}
  are using similar MAC-E-Filters differing somewhat
  in size: The diameter and length of the Mainz (Troitsk) spectrometers are 1\,m
  (1.5\,m) and 4\,m (7\,m). The major differences between the two setups are the tritium
  sources.
\begin{figure}
\epsfxsize=14cm
\centerline{\epsffile{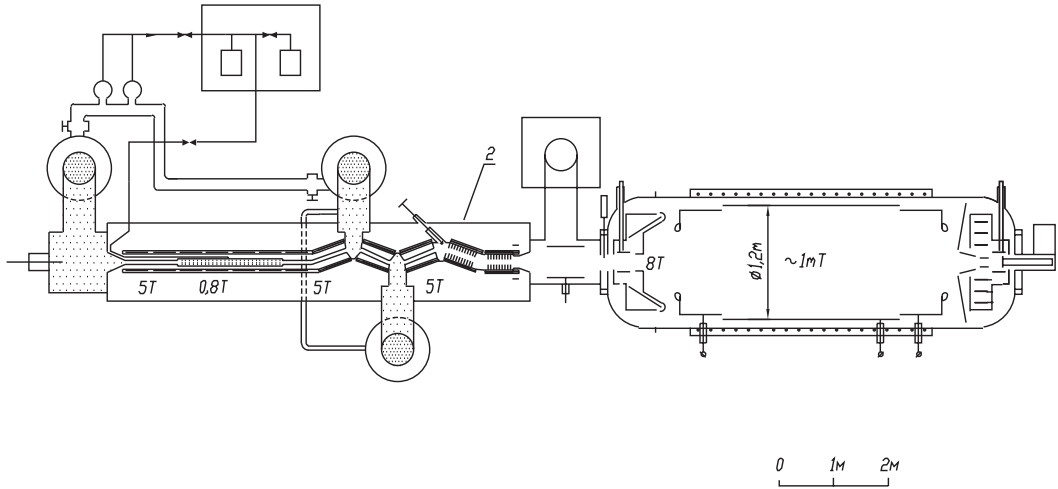}}
  \caption{Schematic view of the Troitsk experimental setup \protect\cite{Belesev95}.}
  \label{trk-setup}
\end{figure}

The Troitsk experiment uses a windowless gaseous tritium source
(WGTS) which is based on the adiabatic transport of electrons in a
strong longitudinal magnetic field and circulation of tritium gas
at low pressure by means of a differential pumping system
\cite{lob85} (see fig.~\ref{trk-setup}). This approach was first
pioneered in an experiment at Los Alamos \cite{LANL}. An essential
refinement made at Troitsk was the use of a strong magnetic field
for electron transport. This technique permits to use multiple
bends of the transport channel, thus providing better differential
pumping and smooth coupling to the MAC-E-Filter spectrometer. The
Troitsk WGTS (a 3\,m long tube of 50\,mm diameter filled with 0.01
mbar of \ttwo ) provides a number of beneficial features for the
study of the tritium $\beta$ spectrum, such as guaranteed
homogeneity over the cross section of the source and reliable
on-line control of inelastic energy losses of electrons in the
source. It further allows to use theoretical calculations of free
molecular final state corrections and almost totally suppresses
back scattering.

Mainz uses a film of molecular tritium quench-condensed onto a
substrate of highly oriented pyrolytic graphite.
  \begin{figure}
  \begin{picture}(14,6)
    \put(8,3){\makebox(0,0){\includegraphics[angle=0,width=14cm]{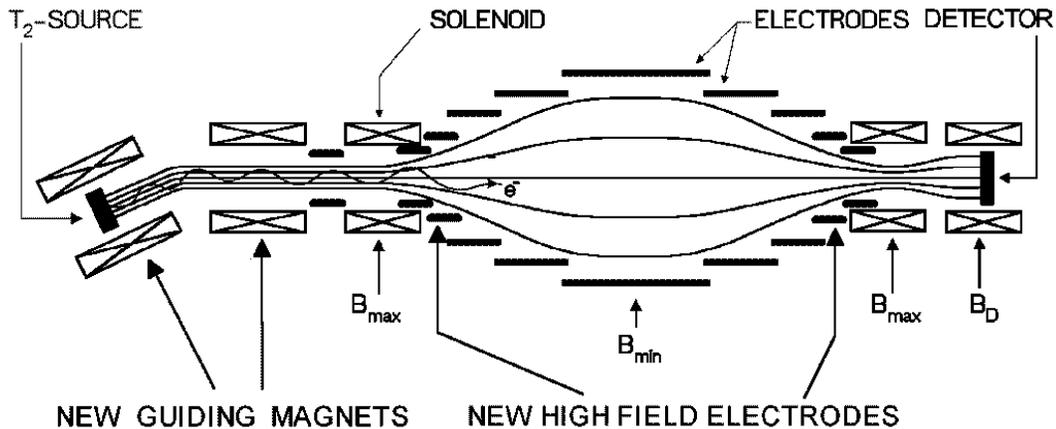}}}
  \end{picture}
  \caption{The upgraded and improved Mainz setup (schematically,
    not in realistic scale). The outer diameter amounts to 1\,m, the distance
    from the source to the detector is 6\,m \protect\cite{Weinheimer99}.}
  \label{fig_newsetup}
  \end{figure}
The film has a diameter of 17\,mm and a typical thickness of 40\,nm, which
is measured by laser ellipsometry. In the years 1995--1997 the Mainz setup was
upgraded to enhance the count rate and to decrease the background rate. As second
substantial improvement a new cryostat now provides temperatures of the tritium film
below 2\,K to avoid a roughening transition of the film, which was a source of
systematic errors of earlier Mainz measurements.
  The roughening process is a temperature activated surface diffusion
  process, therefore low temperatures are necessary to get time constants
  much longer than the duration of the measurement \cite{Fleischmann1,Fleischmann2}.
  The full automation of the apparatus and remote control allows to perform
  long term measurements over several months per year.
  Figure \ref{fig_newsetup} gives a sketch of the Mainz setup.
  Since this upgrade, the count rate, background and energy resolution of the Mainz
setup are about the same as the ones of the Troitsk experiment.

\subsection{Results of the Troitsk Neutrino Mass Experiment}
The Troitsk experiment has taken tritium data for 200 days from
1994 on. Since the first measurements, the Troitsk experiment has
observed a small anomaly in the energy spectrum, located a few eV
below the $\beta$ endpoint \ezero . The distortion resembles sharp
step of the count rate \cite{Belesev95}. Since a MAC-E-Filter is
integrating, a sharp step corresponds to a narrow line in the
primary spectrum. The data indicate a relative intensity of about
$10^{-10}$ of the total decay rate. From 1998 on, the Troitsk
group reported that the position of this line seems to oscillate
with a frequency of 0.5 years between 5\,eV and 15\,eV below
\ezero\ \cite{Lobashev99}. The cause for such an anomaly is not
known. Detailed investigations as well as synchronous measurements
with the Mainz experiment are under way and will help to clarify
this effect.

Fitting a standard \bspec\ to the Troitsk data results in
significantly negative values of $\mnue^2 \approx -10$ to
-20\,eV$^2$. However, describing the anomaly phenomenologically by
adding a monoenergetic line with free amplitude and position to a
standard \bspec\ results in values of \mtwo\ compatible with zero.
The average over all Troitsk runs until 1999 then amounts to
\cite{Lobashev00}
\begin{displaymath}
    \mnue^2  = (\,-1.0 \pm   3.0 \pm 2.5\,)~eV^2
\end{displaymath}
which corresponds to an upper limit of
\begin{displaymath}
  \mnue \leq 2.5 ~eV^2 ~~~~~{\rm (95~\%~C.L.)}
\end{displaymath}

\subsection{Results of the Mainz Neutrino Mass Experiment}


\begin{figure}
\epsfxsize=9.5cm
\centerline{\epsffile{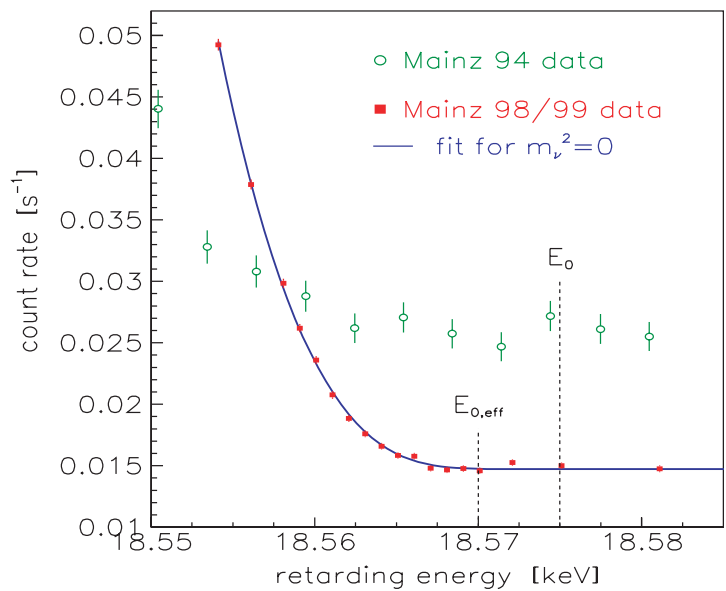}}
\caption{Averaged count rate of the Mainz 1998 and 1999 data
(points) with fit (line) in comparison with previous Mainz data
from 1994 \protect\cite{Bonn97} as function of the retarding
energy near the endpoint \ezero , and effective endpoint $E_{\rm
0,eff}$. The position of the latter takes into account the width
of response function of the setup and the mean rotation-vibration
excitation energy of the electronic ground state of the
$^3$HeT$^+$ daughter molecule.} \label{fig_mainzbetaspec}
\end{figure}
Figure \ref{fig_mainzbetaspec}
  shows the endpoint region of the Mainz 1998 and 1999 data in comparison with the former
Mainz 1994 data. An improvement of the signal-to-background ratio by a factor 10 as well
as a significant enhancement of the statistical quality of the data is clearly visible.
  The main systematic uncertainties of the Mainz experiment
  are originating from the physics and the properties of the quench-condensed
  tritium film and originate from the inelastic scattering of \belec s within the tritium
  film,
  the excitation of neighbor molecules due to the \bdec , and the self-charging of the
  tritium film by its radioactivity. As a result of detailed investigations
  \cite{Aseev,Erice,Bornschein}, these systematic uncertainties were reduced
  significantly.

  The data of the last runs of 1998 and 1999 (see fig. \ref{fig_mainzbetaspec})
  neither show a Troitsk-like anomaly nor any other residual problem.
  The most sensitive analysis on the neutrino mass, in which only the last
  70\,eV of the \bspec\ below the endpoint are used results in
  \begin{displaymath}
  \mnue^2   =  (\,-1.6 \pm 2.5 \pm 2.1\,)~eV^2
  \end{displaymath}
which is compatible with a zero neutrino mass and corresponds to an upper limit
on the electron neutrino mass of:
  \begin{displaymath}
    \mnue \leq 2.2 ~eV ~~~~~{\rm (95~\%~C.L.)} \quad {\rm .}
  \end{displaymath}
The analysis of the new Mainz 1998 and 1999 data \cite{Bonn2000}
improved the published former upper limit of $\mnue<2.8$\,eV
\cite{Weinheimer99}, which was based on the Mainz 1998 data alone.
Together with the Troitsk results, they represent the world's best
sensitivity on a neutrino mass in a direct neutrino mass
experiment.

\section{The KATRIN experiment}\label{chap3}

The tritium \bdec\ experiments at Troitsk and Mainz have almost
reached their limit of sensitivity. It can be estimated that
future measurements of both experiments would improve the current
limit only marginally to $\mnue < 2$\,eV. To measure an electron
neutrino mass in the sub-eV region thus requires a new experiment
with much higher $\nu$-mass sensitivity.

In the following sections we present a design study for a
next-generation tritium \bdec\ experiment with a sensitivity to
sub-eV neutrino masses, following first ideas presented in
\cite{Erice,macetof}. The experiment we propose, the
\underline{Ka}rlsruhe \underline{Tri}tium \underline{N}eutrino
(KATRIN) experiment, would have an estimated sensitivity of $\mnue
= 0.35$\,eV (90\,$\%$ C.L.), which is about one order of magnitude
better than the sensitivity of the current experiments. For
$m^2(\nue)$, which is the observable in a direct neutrino mass
measurement, this corresponds to an improvement by two orders of
magnitude. This requires significant improvements of the tritium
source strength and the spectrometer resolution.

The proposed KATRIN set-up is based on the long-term experience
with the existing spectrometers of the MAC-E type
\cite{lob85,pic92a} and has been prepared by groups from Fulda,
Karlsruhe, Mainz, Prague, Seattle and Troitsk.

\subsection{Experimental overview}

\begin{figure}
\epsfxsize=15cm \centerline{\epsffile{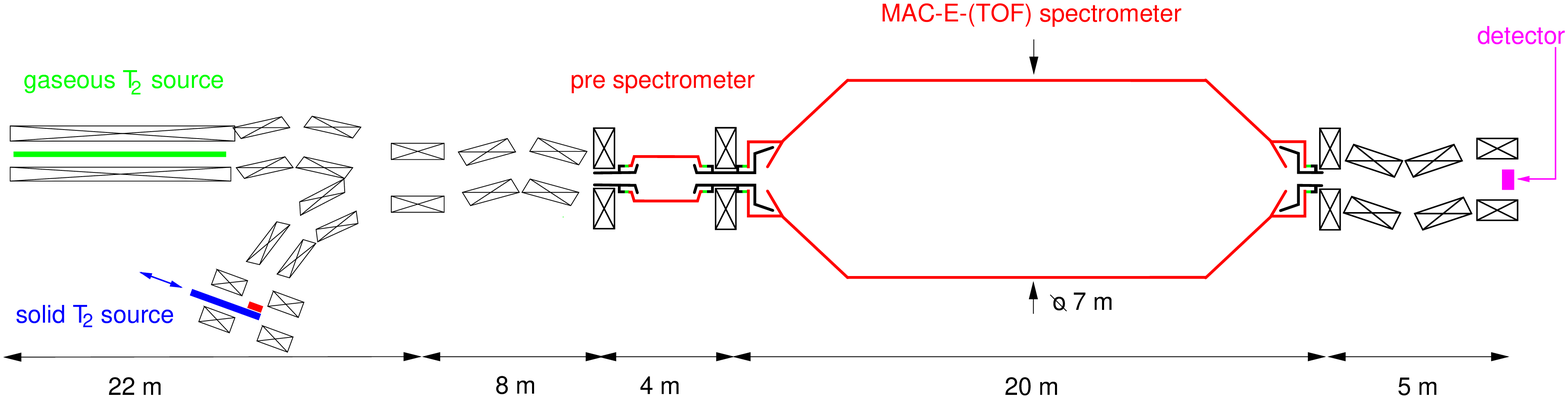}}
\epsfxsize=2.4cm \epsfclipon \centerline{\hspace*{7cm} \small{
Present Mainz Setup\,:} \quad \epsffile{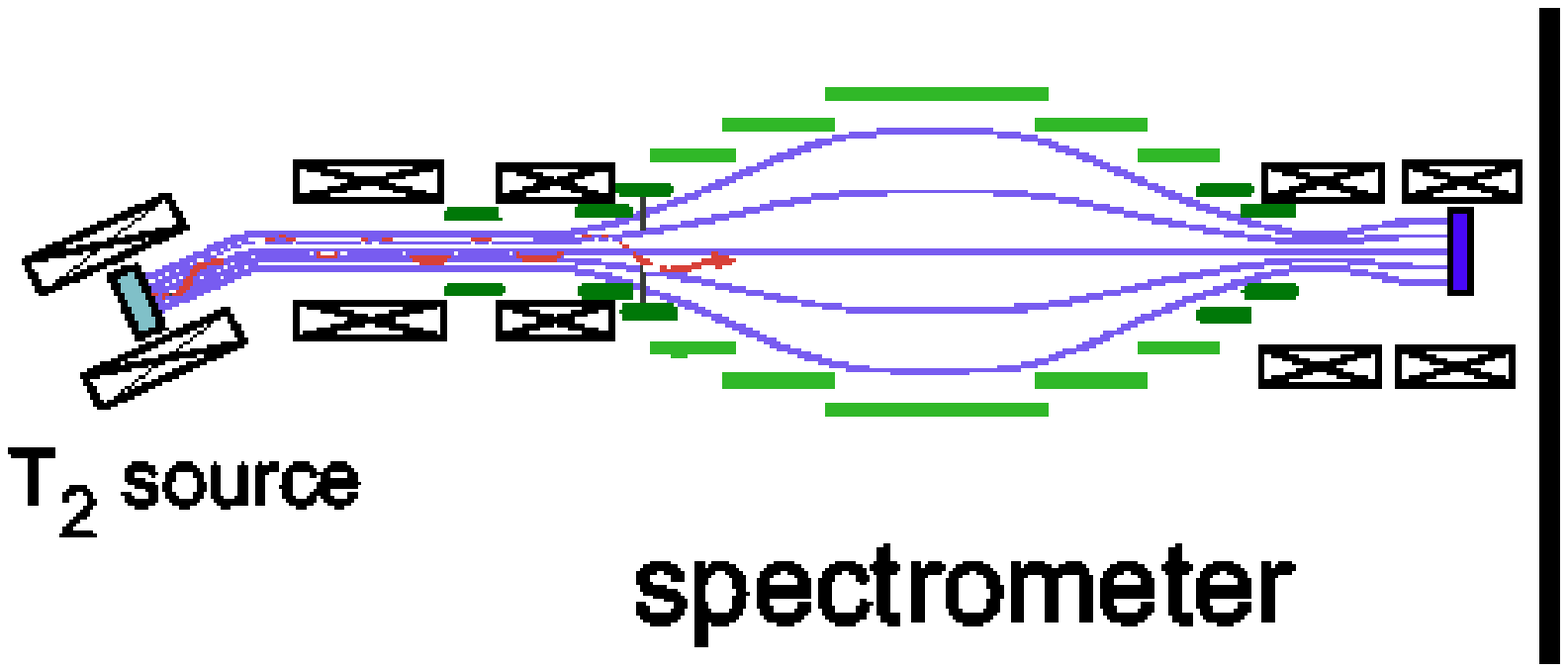}}
\caption{Schematic view of the proposed next-generation tritium
  \bdec\ experiment KATRIN. The main components of the
  system comprise a windowless gaseous tritium source (WGTS), a
  quench condensed tritium source (QCTS), a pre-spectrometer
  and a large electrostatic
  spectrometer of 7\,m diameter and 20\,m length with an energy resolution
  of 1\,eV. An electron transport system guides electrons from the
  \ttwo\ sources to the spectrometers, while eliminating all
  tritium molecules in a two stage process, consisting of a
  differential pumping part followed by a cryotrapping part
  The pre-spectrometer having a retarding potential of about 100~V below
  the $\beta$ endpoint
  allows only the high energy tail of the \belec s,
  comprising  about 2\,$\times$\,$10^{-7}$ of the total decay rate,  to enter
  the main spectrometer.
  The overall length of this linear set-up amounts
  to about 70\,m. Shown for comparison is the present Mainz setup
  at the same scale.} \label{lin_setup}
\end{figure}

 A schematic view of the proposed experimental configuration is shown in
 fig. \ref{lin_setup}. It can be grouped into four functional units:
 \begin{itemize}
 \item the two molecular tritium sources WGTS and QCTS (including tritium supply and handling)
 \item the electron transport and tritium elimination line
       (comprising the differential pumping and cryotrapping sections)
 \item the electrostatic pre- and main-spectrometer
 \item the $\beta$ electron detector
 \end{itemize}
In order to reach a sub-eV sensitivity, an energy resolution of
$\Delta E=1$ eV  is necessary at the tritium \bdec\ endpoint of
18.6\,keV, implying a ratio of magnetic fields \bmin\,/\,\bmax = 5
$\times$ 10$^{-5}$. This resolution would correspond to an
improvement of a {\it factor of 4} compared to the experiments in
Troitsk and Mainz. Since the energy interval of interest below the
endpoint $E_0$ (see fig. \ref{fig_betaspec}) rapidly decreases
with smaller neutrino mass $m_\nu$, the signal rate has to be
increased. This can be achieved by a higher \ttwo\ source strength
corresponding to a larger source area \as\ and a higher and
optimized column density $\rho d$. Subsequently, this requires a
larger area of the analyzing plane \aan\ of the spectrometer. To
meet these demands, a large electrostatic spectrometer with an
analyzing plane of 7\,m diameter and a T$_2$ source with effective
cross section of $\aseff \approx 16$~cm$^2$ and column density of
$(\rho d)_{\rm eff} \approx 5\cdot 10^{17}$ molecules/cm$^2$ are
considered for KATRIN.

In addition, a low background count rate of the order of 10$^{-2}$
counts/s or less at the tritium $\beta$-decay endpoint region at
18.6 keV is  required when looking for sub-eV neutrino mass
effects. Among the various background processes, the ionization of
residual gas molecules as well as the decay of any residual
tritium in the spectrometer play an important role. Therefore, the
low background count rate is implying stringent requirements on
the vacuum conditions of the electrostatic spectrometers. To
further reduce the background in the main spectrometer, a
pre-spectrometer (located between the tritium source and the
main-spectrometer) will act as a pre-filter at a retarding energy
of order of 100~eV below the $\beta$ endpoint $E_0$, reducing the
amount of \belec s entering the main-spectrometer by about 7
orders of magnitude. Suppression of cosmogenic background finally
calls for a state-of-the-art detector with good energy- and
spatial resolution for low energy electrons in the keV range.

During the longterm tritium measurements special emphasis will be
put on the control of all systematic effects which might influence
the experimental results. As the most
 important systematic effects are associated with the properties of
 the tritium source, two independent tritium sources with different
 systematics are proposed for KATRIN: a windowless
 gaseous tritium source (WGTS), following the design of the Troitsk
 experiment, and a quench condensed tritium source (QCTS),
 following the Mainz design. Alternate measurements with both
 sources will minimize systematic uncertainties. Both sources will
 have to be calibrated and controlled extensively. Calibration of
 the properties of the QCTS will be provided by a series of
 $^{83m}$Kr-conversion line measurements, while the characteristics
 of the WGTS will be determined by means of an electron gun and of
 $^{83m}$Kr.

 Apart from the standard integrating (MAC-E) mode of operation,
 short-term measurements with the differential time-of-flight
 (MAC-E-TOF) mode \cite{macetof} will play an
 important role in the measuring programme of KATRIN.
 These additional runs will help to investigate
 systematic uncertainties -- {\it e.g.} inelastic scattering
 of the \belec s in the tritium source -- with much higher
 precision than the integral MAC-E-Filter mode would allow.

 The detailed spectral information obtained in the MAC-E-TOF mode
 will also allow to search for non-SM physics like possible small
 right-handed contributions to the electroweak interactions \cite{stephenson},
 or tachyonic neutrinos \cite{ciborowski}. Also, effects such as the anomaly
 reported in the Troitsk experiment \cite{Belesev95} can be identified and
 investigated in the non-integrating MAC-E-TOF mode more clearly.
 Moreover, the resulting distortion in the energy spectrum will stay {\it local}
 in a non-integrating spectrometer mode.
 This in turn will allow to disentangle the
 effects of non-zero $\mnue^2$ values
 from any narrow spectral anomaly close to
 the endpoint.

 The favorable location for a future high precision tritium \bdec\
 experiment is the site of Forschungszentrum Karlsruhe (FZK),
 which offers general infrastructure matching well the extensive
 experimental demands. In particular, a location on site of FZK
 offers close proximity to a tritium laboratory (Tritium Labor
 Karlsruhe, TLK), which is well
 suited to handle and to process the total tritium inventory of the
 experiment of about $10^{13}$\,Bq. The hall to house the KATRIN experiment
 has to be located sufficiently close to the tritium
 laboratory TLK to allow the continuous supply of
 tritium gas to the experiment via transfer lines (feed and return tubes).
 For a linear set-up of all components, as shown in the schematic view of
 fig. \ref{lin_setup}, the overall length of the KATRIN hall
 would be about 70\,m. Therefore we propose to
 set-up and to operate the experiment on the site of FZK.

 In the following sections we discuss the experimental parameters
 relevant for the improvement in statistics and resolution, namely the two
 tritium sources, the electron transport sections, the two spectrometers
 and the detector options. This is followed by a discussion of the background
 and of the systematic uncertainties. Finally, we point out the physics objectives of
 the KATRIN experiment, in particular we present sensitivity estimates
 for the electron neutrino mass.

\subsection{Experimental parameters}
\label{sec_parameters}

For tritium $\beta$ decay experiments like KATRIN, which are based
on the MAC-E- Filter technique, the ratio of magnetic field
strengths in different parts of the experiment (tritium source,
magnetic pinch, analyzing plane of the spectrometer and detector)
determines several key experimental parameters.

\subsubsection{Magnetic fields}

\begin{figure}
\epsfxsize=11cm \centerline{\epsffile{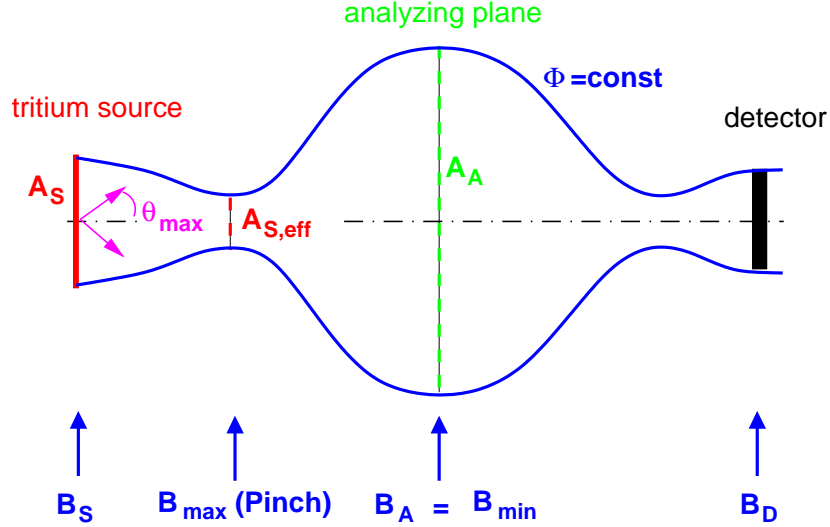}}
\caption{Schematic view of magnetic fields and cross sections
(omitting the pre-spectrometer). The magnet field values are $\bso
 = 6$~T, $\bmax = 10$~T,
 $\ba = \bmin = 5\,\times 10^{-4}$~T and $\bd = 3$~T (\,all values
 are subject to a common scaling factor $f\,=\,0.5-1.0$\,).
 }
\label{fig_bfields}
\end{figure}

The geometry and the arrangement of magnetic fields is illustrated
schematically in fig. \ref{fig_bfields}.
The tritium source with area \as\ is placed in a magnetic field
$B_S$, which is lower than the maximum magnetic field \bmax\
(magnetic pinch). Their ratio determines the maximum accepted
starting angle $\thmax = \arcsin \sqrt{\bso / \bmax }$ (see eq.
\ref{eq_thetamax}). Due to the conservation of the magnetic flux
$\Phi$ within which the electrons are transported, an effective
source area \aseff\ at the magnetic pinch can be defined. The
maximum cross section \aan\ of the flux tube is reached in the
analyzing plane  at the minimum magnetic field $\ba$\,. The ratio
\ba/\bmax\ determines the relative energy resolution according to
eq. (\ref{eq_resolution}).

The following considerations are based on the assumption of a
conserved magnetic flux $\Phi= B_S \cdot \as = \bmax \cdot \aseff
= \ba \cdot \aan $ of
\begin{equation}
  \Phi= f \cdot 5\cdot 10^{-4}\,{\rm T}\cdot \pi (7/2\,{\rm m})^2
      = f \cdot 192 \,{\rm Tcm}^2
\label{eq_magflux}
\end{equation}
The common scaling factor $f$ with  $0.5 \leq f \leq 1.0$ allows
for further adjustments of magnetic fields to their final values.
The current values (see fig. \ref{fig_bfields}) have been obtained
by optimizing the requirements for adiabaticity, which aims for
high magnetic fields. Avoiding the trapping of particles, on the
other hand, would favor lower magnetic fields. In addition,
technical feasibility and costs were considered.

\subsubsection{Transmission and response function}
The transmission function {\it T} of a MAC-E-Filter is fully
analytical and depends only on the two field ratios $\ba / \bmax$
and $\bso / \ba$\,:
\begin{equation}
  \label{eq_mace}
  T(E,qU_0)
     = \left\{ \begin{array}{l@{~~~~~}l}
                   0 & E-qU_0 < 0\\
                   \frac{1 - \sqrt{1 - \frac{E-qU_0}{E} \cdot
                                   \frac{B_{\rm S}}{B_{\rm A}}}  }
                   { 1 - \sqrt{1 - \frac{\Delta E}{E}
                                   \cdot \frac{B_{\rm S}}{B_{\rm A}}} }
                      & 0 \leq E-qU_0 \leq \Delta E\\
                   1 & E-qU_0 > \Delta E
                \end{array} \right.
\end{equation}

with E denoting the electron energy and $qU_0$ defining the
retarding energy. The total width $\Delta E$ of the transmission
function from $T=0$ to $T=1$ is given by $\ba / \bmax$ (see
\ref{eq_resolution}). The shape of {\it T} in this interval is
determined by $\bso / \ba$, as this ratio defines the maximum
accepted electron starting angle \thmax\ (see eq.
\ref{eq_thetamax}).

\begin{figure}[tb]
\epsfxsize=10cm \centerline{\epsffile{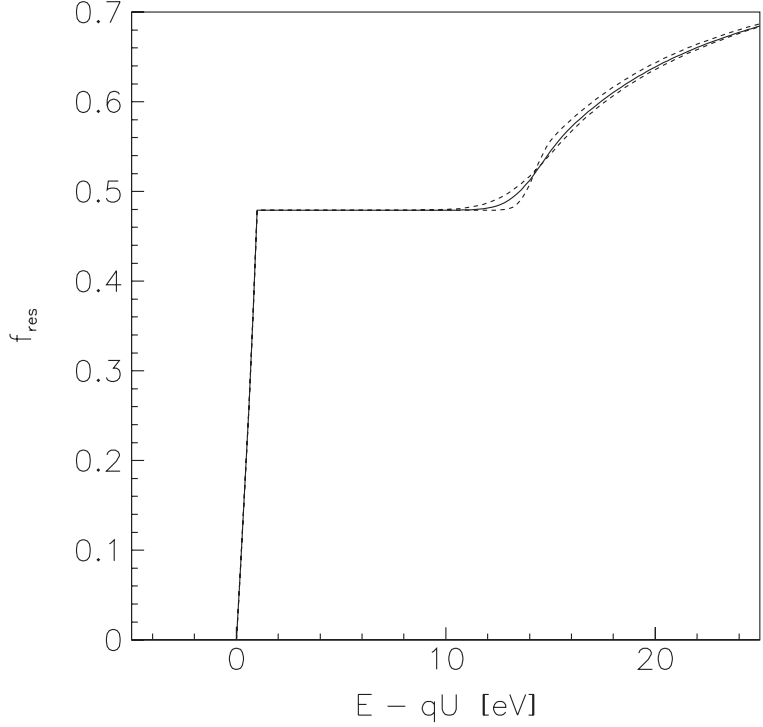}}
\caption{Response function of the KATRIN spectrometer for
  isotropically emitted electrons with fixed energy $E$ as a function of
  the retarding energy $qU$. The energy loss of electrons in the WGTS source
  (column density $\rho d$ = 5 $\times$ 10$^{17}$\,/\,cm$^2$, maximum starting
  angle $\theta_{max}$ = 51 $\deg$) is folded into the response function.
  To illustrate the uncertainties of the
  shape of the energy loss function, the experimentally derived
  function (solid line) and its uncertainties (dashed lines) derived
  for quench condensed tritium
  according to reference \protect\cite{Aseev} were used.}
  \label{fig_response}
\end{figure}

The transmission function {\it T} does not take into account the
interactions of electrons in the source. Electrons which have
undergone inelastic collisions with \ttwo\ molecules in the source
have suffered an energy loss and hence have a modified
transmission function. These processes can be described by folding
of the corresponding inelastic cross section \cite{Aseev} with the
distribution of electron path lengths through the source
(including multiple scattering). Folding this energy loss
distribution of electrons in the source with the transmission
function (\ref{eq_mace}) defines the so-called response function
$f_{res}$ of the experiment.

In fig. \ref{fig_response} the response function $f_{res}$ of
KATRIN is shown for isotropically emitted monoenergetic particles
(with fixed energy $E$) as a function of the retarding energy
$qU$. The figure is based on the following given standard
parameters: a) energy resolution $\Delta E = 1$~eV, b) WGTS column
density $\rho d = 5 \cdot 10^{17}/$cm$^2$ and  c) maximum accepted
starting angle $\thmax = 51^\circ$. The influence of the electric
potential drop and the inhomogeneity of the magnetic field in the
analyzing plane has not been considered, since these effects can
be compensated by a detector with a radial position sensitivity.

Due to the sharpness of $\Delta E$ of KATRIN and due to the high
threshold of the \ttwo\ excitation, the ``no energy loss''
fraction of transmitted electrons can clearly be separated from
those electrons which have undergone inelastic collisions (see
fig. \ref{fig_response}). The former fraction corresponds to the
sharp rise from 0 to the flat plateau, whereas the latter part
represents the second upslope at the abscissa of about 12~eV. The
relative amounts of each fraction ({\it i.e.} the relative height
of the elastic plateau in fig. \ref{fig_response}) is determined
by the column density of the source.

For the case of the continuous $\beta$ electron spectrum from
tritium decay, the $f_{res}$ function of KATRIN implies that the
last ~10~eV below the endpoint \ezero\ are fully covered by the
elastic plateau of the response function. This means that the
region with the highest sensitivity to \mnu\ is not affected by
inelastic processes. Even with a larger measuring interval of
25~eV below \ezero\,, inelastic events contribute to only 2~\% of
the signal rate (compare fig. \ref{fig_response}). Therefore, the
systematic uncertainties due to energy losses within the source
become dominant only in the analysis of large energy intervals
(see fig.~\ref{fig_sensitivity}).

\subsubsection{Signal rates}
The counting statistics is defined by 3 parameters: the signal
rate $S$, the background rate $B$ and the total measurement time
$t_{tot}$.

For a tritium source of size \as\ and column density $\rho d$ and
maximum starting angle \thmax\ accepted by the spectrometer, the
signal count rate $S$ very close to the endpoint is proportional
to the number of tritium molecules $N(T_2)~\propto \as \cdot \rho
d$, the relative accepted forward solid angle $\Delta \Omega /2\pi
= 1 - \cos \thmax $, and the probability of a $\beta$ electron not
to undergo an inelastic scattering process $P_0(\rho d,\thmax
)$\footnote{Any inelastic scattering process requires a minimum
amount of energy loss of 10 eV (s. fig. \ref{fig_response}),
therefore only the zero loss fraction P$_0$ contributes to the
count rate very close to the endpoint. P$_0$ is an average over
all paths of electrons, starting angles up to \thmax \ and
starting points within the source column density.}:
\begin{eqnarray}
  S & \propto & N(T_2) \cdot \frac{\Delta \Omega}{2 \pi}
      \cdot P_0(\rho d,\thmax )
      \label{eq_signal}\\
    & = & \as \cdot \rho d \cdot (1- \cos \thmax )
      \cdot P_0(\rho d,\thmax ) \label{eq_s_with_as}\\
    & = & \frac{\aan \cdot \Delta E}{E} \cdot
          \frac{1}{1+\cos \thmax }
                      \cdot \rho d \cdot P_0(\rho d,\thmax )
          \label{eq_s_with_deltae}\\
    & = & \frac{\aan \cdot \Delta E}{E} \cdot (\rho d)_{\rm eff}
          \label{eq_s_with_rhodeff} \\
    & \leq &
      \frac{\aan \cdot \Delta E}{E} \cdot \frac{(\rho d)_{\rm free}}{2}
           \label{eq_s_with_lambdafree}
\end{eqnarray}
Equation (\ref{eq_s_with_deltae}) follows through eq.
(\ref{eq_s_with_as}) from magnetic flux conservation $B \cdot A =
{\rm const.}$ and from eq. (\ref{eq_resolution}) and
(\ref{eq_thetamax}). Restricting the signal rate to the elastic
fraction only is possible due to the narrow energy interval of the
KATRIN measurements(as discussed above). Thus the last three terms
of eq. (\ref{eq_s_with_deltae}) can be understood as an effective
column density $(\rho d)_{\rm eff}$ of a virtual source of
electrons not undergoing any scattering process and placed in the
maximum magnetic field \bmax\ and emitting into the full forward
solid angle of $2 \pi$ from an effective source area $\aseff =
\aan \Delta E/E$. The signal rate $S$ of eq.
(\ref{eq_s_with_deltae}) increases for all \thmax\ with larger
column density $\rho d$, but since $P_0(\rho d,\thmax )$ decreases
at the same time, the effective column density $(\rho d)_{\rm
eff}$ approaches a maximum asymptotic value (see also
fig.~\ref{fig_enloss}):
\begin{equation}
 (\rho d)_{\rm eff} =
  \frac{1}{1+\cos \thmax } \cdot \rho d \cdot P_0(\rho d,\thmax )
  \stackrel{d \rightarrow \infty}{\rightarrow} \frac{(\rho d)_{\rm free}}{2}
   = \frac{1}{2 \cdot \sigma}
  \label{eq_limit_largerhod}
\end{equation}

Equation (\ref{eq_limit_largerhod}) means that averaging over all
emitted starting angles and weighting with the accepted solid
angle the maximum effective source thickness is restricted to half
of the mean free column density $(\rho d)_{free} = 1/\sigma =
(2.94 \pm 0.06) \cdot 10^{17} ~$cm$^{-2}$, where $\sigma$ is the
total inelastic cross section at 18.6~keV \cite{Aseev}.

Equation (\ref{eq_s_with_lambdafree}) shows that the maximum
signal rate, which can be achieved is determined by the product of
two key experimental parameters: the energy resolution $\Delta E$
and the size of the analyzing plane \aan\ of the electrostatic
spectrometer. For KATRIN the gain in \aan\ by a factor of about
100 will in parts be counteracted by the improvement of $\Delta E$
by a factor 4. On the other hand, the restriction to the elastic
fraction will allow to increase the column density substantially
with respect to the presently used value of $2.4 \cdot
10^{17}\,{\rm molecules/cm}^2$ \cite{Lobashev99} without
increasing the systematic uncertainties.

\subsection{Windowless gaseous tritium source}

\begin{figure}
\epsfxsize=9cm \centerline{\epsffile{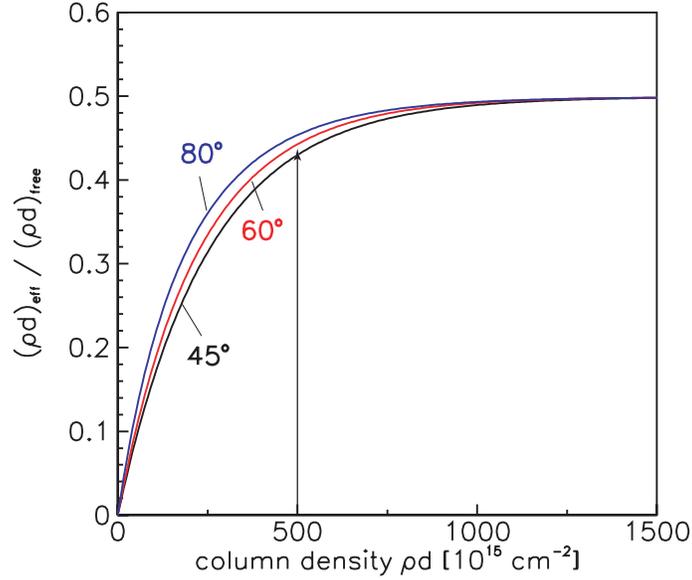}}
\caption{Ratio of effective to free column density
  $(\rho d)_{\rm eff} / (\rho d)_{\rm free}$
  (being proportional to the signal rate $S$) as function of the source column
  density $\rho d$  for different
  maximum accepted starting angles \thmax . The vertical
  line indicates the proposed WGTS parameters of
  $\rho d = 5 \cdot 10^{17}\,{\rm molecules/cm}^2$ and $\thmax =
  51^{\circ}$.
  } \label{fig_enloss}
\end{figure}

 The windowless gaseous tritium source (WGTS) allows the measurement of the
 endpoint region of the tritium \bdec\ and consequently  the determination
 of the neutrino mass with a minimum of systematic uncertainties from the tritium
 source. Such a source was first used at the LANL experiment \cite{LANL}
 and developed further and
 adapted to the MAC-E-Filter by the Troitsk group \cite{lob85}.

Figure~\ref{fig_enloss} can be used to obtain realistic design
parameters of the WGTS for KATRIN. The graph shows the ratio of
effective to free column density (compare eq.
(\ref{eq_s_with_rhodeff}), (\ref{eq_s_with_lambdafree})) as a
function of the real column density of the source. Choosing the
latter as $5 \cdot 10^{17}\,{\rm molecules/cm}^2$ yields a ratio
which is reasonably close to the asymptotic maximum for all
\thmax\,. With respect to the present Troitsk source the gain in
$(\rho d)_{\rm eff}$ would be a factor of about 1.5. Together with
the proposed gain in \aan\ by a factor of 100 and the reduction of
$\Delta E$ by a factor of 4 the gain factor in signal rate close
to the endpoint would be about 40.

\begin{figure}
\epsfxsize=14cm \centerline{\epsffile{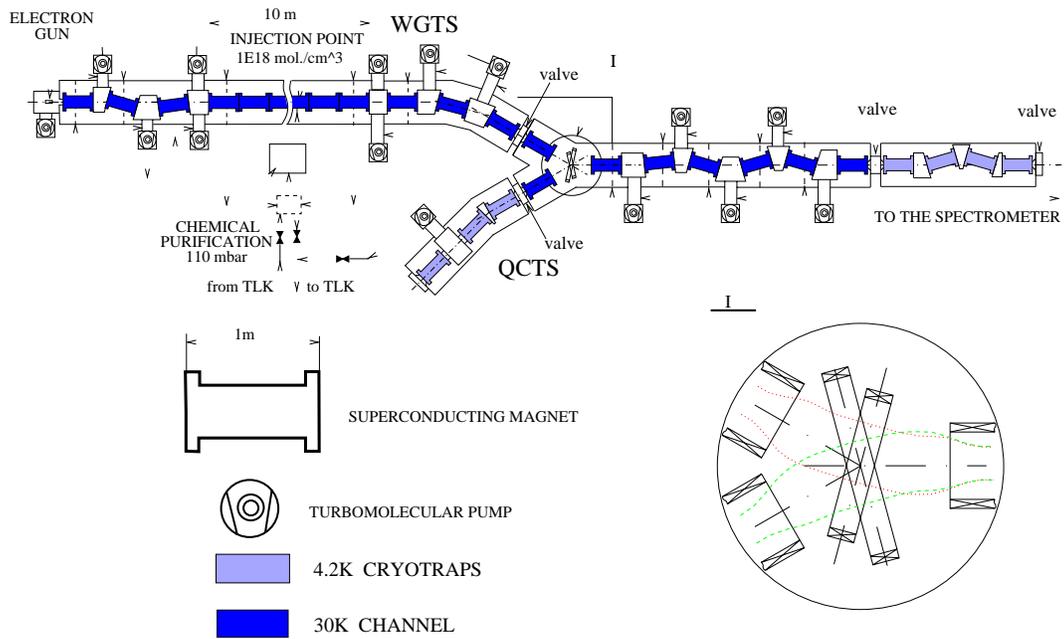}}
\caption{Schematic view of the windowless gaseous tritium source
(WGTS) and of the electron transport and differential pumping
system. This system includes the y-shaped solenoidal beam switch
(also shown enlarged in the inset). Energizing only one of the two
short coils and only two of the three transport solenoids allows
to select the direction of the electron beam and thus to change
the tritium source. The dotted lines in the inset represent the
magnetic field lines. } \label{fig_wgts}
\end{figure}

The WGTS will consist of a 10\,m long cylindrical tube of 70\,mm
diameter (see fig.~\ref{fig_wgts}), filled with molecular tritium
gas of high isotopic purity ($> 95$~\%). The tritium gas will be
injected by a capillary at the middle of the tube forming there a
density of $10^{15}$ molecules/cm$^3$. It then diffuses over a
length of 5\,m to both end faces of the tube, resulting in a
linear decrease of tritium density by a factor of 100 from the
injection point to the end faces. In order to keep the source
strength constant, the tube and the injected tritium gas have to
be temperature stabilized to $\pm$ 0.2 degrees. A working
temperature around 30\,K optimizes the WGTS column density. Then
the central \ttwo\ pressure is $4 \cdot 10^{-3}$~mbar. Cooling is
achieved by means of circulating He gas of the appropriate
temperature in a small pipe thermally coupled to the tritium tube.

It is proposed to place the tritium tube inside a chain of ten
superconducting solenoids of 1\,m length each. The solenoids will
generate a homogeneous magnetic field of $B_S = 6$\,T, which
adiabatically guides the decay electrons to the end faces (the
modular design of the solenoids has been chosen for reasons of
quench stability). The ratio of magnetic field strengths at the
source  and at the pinch magnet $\bso / \bmax$ is chosen to be
0.6, so that the maximum accepted starting angle in the WGTS is
$\thmax = 51^{\circ}$ and the accepted source area is $\as =
32$~cm$^2$ (compare eq. (\ref{eq_thetamax}) and
(\ref{eq_magflux})).

The tritium supply for the WGTS will be provided by four
double-walled transfer lines from the TLK (one feed and three
return lines for different contamination levels). TLK has worked
out a procedure to guarantee the supply of tritium with isotopic
purity better than 99\,$\%$ and a negligible content of other
gases ($^3$He, $^4$He). The purity of the supplied tritium will be
measured by means of gas chromatography. After processing of the
returned gas (detritiation and tritium enrichment) in dedicated
glove boxes, the purified tritium will be recirculated to the
WGTS.

The main advantages of the WGTS can be summed up as follows:
\begin{itemize}
 \item investigation of  the tritium $\beta$-spectrum with the highest
  possible energy
  resolution, limited only by the spectrum of final state vibrational and
  rotational excitations of the daughter molecule ($^3$HeT)$^+$
 \item use of a maximum specific activity (high signal rate)
 \item no perturbing solid state effects (the most serious being
  self-charging of tritium films \cite{Erice})
 \item perfect homogeneity over the source cross section
\end{itemize}
Attention has to be paid to the following points:
\begin{itemize}
 \item stability of source strength
 \item magnetic trapping of charged particles in the local magnetic field
    minima between the solenoids of the source and
    the subsequent differential pumping system.
\end{itemize}

 \subsection{Quench condensed tritium source}
 \label{sec_qcts}

\begin{figure}
\epsfxsize=10cm \centerline{\epsffile{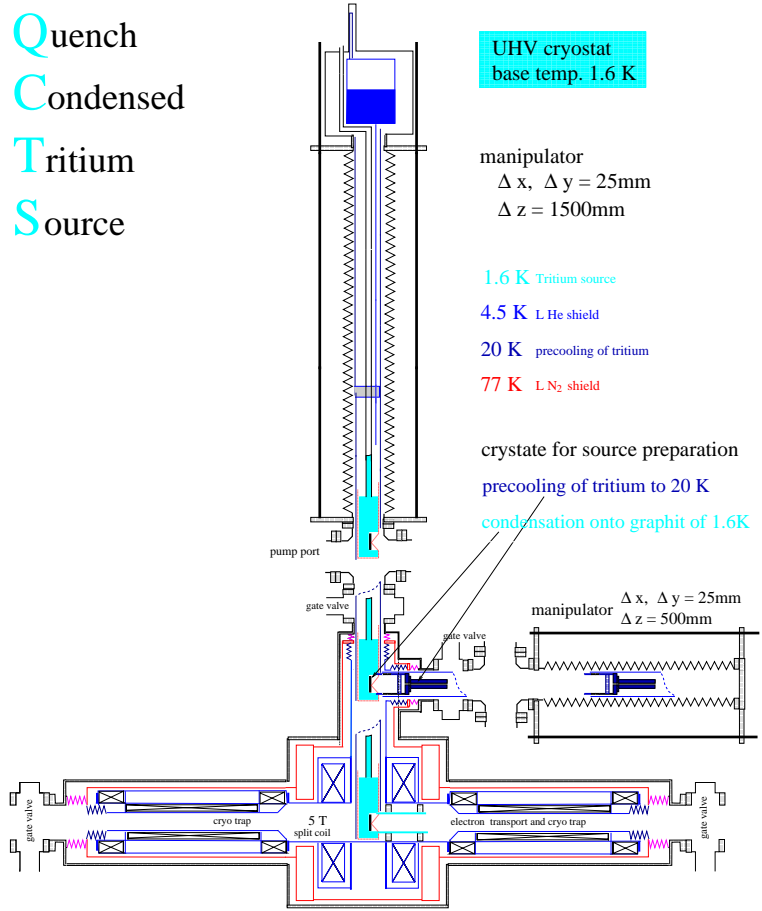}}
\caption{Schematic view of the quench condensed tritium source
(QCTS).} \label{qcts}
\end{figure}

 The quench condensed tritium source (QCTS) will run under rather different
conditions as compared to
 the gaseous WGTS source. The design of the QCTS (see fig.
 \ref{qcts}) largely follows the source concept of the Mainz experiment
 of a thin film of molecular tritium quench condensed on a graphite
 substrate.

 The QCTS will be mounted on a cold-head of a vertical continuous flow
 cryostat with 1.6\,K base temperature and with total height of
 4.5\,m. There are three different positions intended for the QCTS
 source:
 \begin{itemize}
 \item the {\it measuring position} inside a 5\,T split coil magnet
  of 200\,mm diameter and LHe cold bore.
 \item the {\it source preparation area} facing a second continuous flow
  cryostat. This area is equipped with a facility to pre-cool the
  tritium to 20\,K before quench condensing.
  It contains a laser system for surface treatment
  (surface cleaning will be done in-situ by laser ablation) as well as
  standard diagnostics instruments, like a laser ellipsometer for
  film thickness measurements.
 \item the {\it service position} for maintenance of the QCTS at room
  temperature.
 \end{itemize}
 Figure \ref{qcts} shows a possible version
to connect a vertical cryostat to the
 horizontal electron beam line by the help of a high field split coil solenoid.
 Between the QCTS and the y-shape solenoidal beam switch
 a cryotrapping section  is inserted, to avoid
 condensing of residual gas molecules on top of the condensed
 tritium film and to reduce the heat load on the QCTS cryostat.

 The solid source, which is quench condensed onto a highly
 oriented pyrolytic graphite crystal, will be working at
 fixed operating temperature of about 1.6\,-\,1.8\,K.
 This low temperature is required to suppress
 surface diffusion and film roughening \cite{Fleischmann1,Fleischmann2}.
The thickness
 and thus also the luminosity of the QCTS will be limited due to
 self charging \cite{Erice,bornschein2001} to a value of about 100 monolayers
 (1\, monolayer is equivalent to a film thickness of about 3.4\,\AA\ and
 corresponds to a column density
 of the gaseous source of $0.9 \cdot 10^{15}$\,molecules/cm$^2$)
 \cite{Erice,Bornschein}.
 The self-charging generates a linear drop of the electric
 potential from the first to the last film layer of about
 20~mV per monolayer. Thus the optimum energy resolution
 of the main spectrometer for QCTS measurements will be around
 $\Delta E = 1.6$~eV, which corresponds
 to a magnetic pinch field of $\bmax = f \cdot 6$~T.
 Due to its lower column density, the QCTS will be run at a higher
 magnetic field of $B_S = 0.97 \cdot \bmax $, resulting
 in an maximum accepted starting angle of $\thmax = 80^\circ$ and
 a source diameter of about 7\,cm according to eq. (\ref{eq_magflux}).
The corresponding gain in source emittance largely compensates the
loss in column density.
 In total the self-charging effect and the reduced energy resolution will
 limit the {\it effective} energy resolution of
 the frozen source to about 2.5\,eV. Compared to the
 present Mainz source with an area of 2\,cm$^2$ and a thickness of
 140\,ML  the new design would correspond to an
 increase of the count rate by a factor of 30.\footnote{Both the resolution
 and the count rate of the QCTS could be improved significantly, if attempts to suppress the
 self-charging effect turn out to be successful. It is planned to
 explore this in side experiments in collaboration with Prof. Leiderer in
 Konstanz. One possibility could be to inject electrons into the source
 by photo effect at the substrate surface.}

 The effective lifetime of the new QCTS will be determined by the
 rate of tritium losses caused by $^3$HeT$^+$ recoil. This has been found
 to
 be of the order of 0.16\, monolayer per day \cite{Bonn2000}.
 The evaporating tritium  is trapped by the LHe cold cryotraps between the
 QCTS and the beam switch.

 As will be discussed later in section \ref{sec_systematics},
 the QCTS will provide important results with {\it independent} systematic
 uncertainties.

 \subsection{Differential pumping and electron transport system}

 The electron transport system adiabatically guides \bdec\
 electrons from the tritium sources to the spectrometer,
 while at the same time eliminating any tritium flow towards
 the spectrometer, which has to be
 kept practically free of tritium for background and safety reasons.
 The tritium flow (HT molecules) into the spectrometer should be smaller than
 $10^{-13}$~mbar~l/s~\^{=}~$2.7 \cdot 10^6$~molecules/s
 to limit the increase of background to
 $10^{-3}$~counts/s.

\begin{figure}
\epsfxsize=12cm \centerline{\epsffile{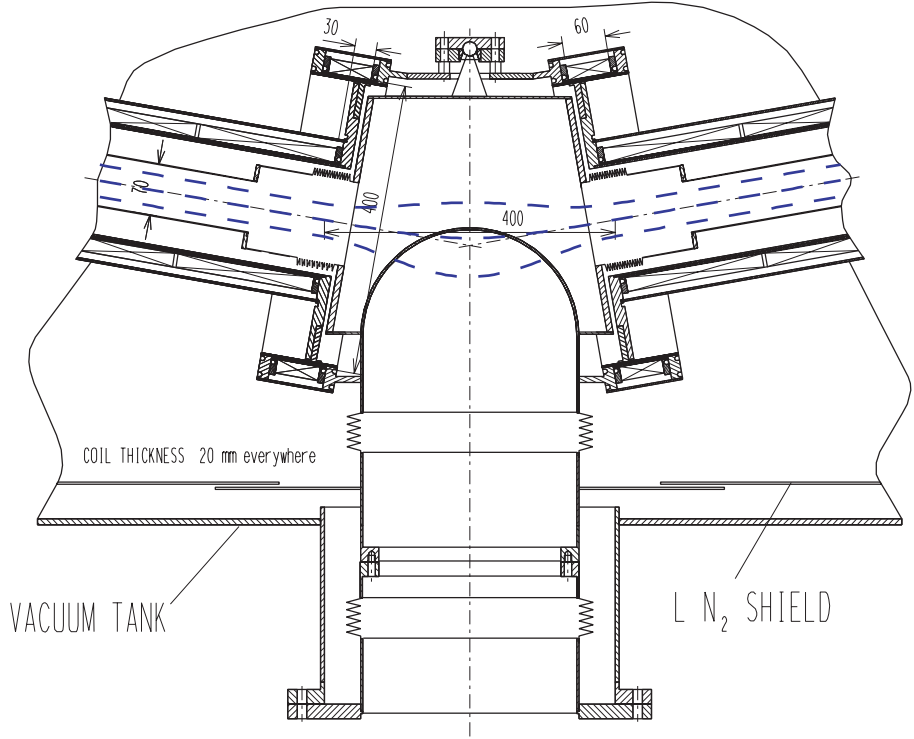}}
\caption{Schematic view of a pumping port between two transport
 magnets. The central magnetic field line as well the ones describing
 the envelope of the electron transporting magnetic flux tube
 (see eq. \ref{eq_magflux}) are illustrated by dashed lines.}
 \label{difpump}
\end{figure}

The first part of the transport system consists of a differential
pumping section directly following the WGTS and consists of 8
transport and pumping stages (see fig.~\ref{difpump}). Each
transport element consists of a 1\,m long tube of 70\,mm diameter
which is kept at a temperature of 30\,K and which is placed inside
a superconducting solenoid ($B=6$\,T). The pumping ports are
located at the gaps between the transport solenoids, which are
tilted by $20^{\circ}$ with respect to each other. The field in
the port is provided by a pair of split coils. The ports will be
equipped with turbo-molecular pumps with a nominal pumping speed
of 2000~l/s. The effective pumping speed of the port is estimated
to be $s_{\rm eff} > 700$~l/s. The conductance of the transport
tube for \ttwo\ at 30~K is 35~l/s. Hence we estimate an extinction
factor of $\geq$ 20 per stage. The beaming effect due to the long
tube changes this result only by 20~\% at maximum. The series of 8
differential pumping stages yields an overall tritium extinction
factor of $\geq\ 2 \times 10^{10}$. Assuming that not more than
one pump may fail at any time, the extinction factor will still be
10$^9$. This value is used as the basis for further calculations.
The flow out of the \ttwo\ source amounts to $10^{19}$~molecules/s
at each end face and hence less than $10^{10}$~molecules/s at the
end of the differential pumping chain. In steady state the \ttwo\
gas may be recirculated directly from the pumps to the source many
times before the isotopic impurity level reaches the margin of
5~\%. Therefore, the \ttwo\ flow from and  back to TLK necessary
for maintaining the required purity can be kept far below the
before mentioned level of $10^{19}$~molecules/s. The second
tritium source of the experiment, the quench condensed QCTS is
connected to the electron transport system by a y-shape solenoidal
beam switch (see fig. \ref{fig_wgts}).

 In the next part of the transport section, the
 cryotrapping section, all remaining traces of tritium will be
trapped onto the liquid helium cold surface of a transport tube
surrounded by transport magnets. It will be covered either by a
thin layer of charcoal or
 argon snow for better trapping.
 Again, 4 individual transport elements of 1\,m
 length and 70\,mm diameter are tilted by $20^{\circ}$ to each other,
thus prohibiting a
 direct line of sight.
The trap will accumulate less than $10^{15}$~molecules/day which is negligible
in view of its huge capacity. Under normal conditions its leakage into the
spectrometers should be essentially zero. Safety valves will protect the latter
in case of failure, {\it e.g.} a warmup.

 \subsection{Electrostatic pre-spectrometer}

 Between the tritium sources and the main spectrometer
 a pre-spectrometer of MAC-E-Filter type will be inserted, acting
as an energy pre-filter to reject all \belec s except the
 ones in the region of interest close to the endpoint \ezero .
 For example, working at a retarding energy 100~eV below
 \ezero\,, only a fraction of $2 \times 10^{-7}$ of the total flux of $\beta$
particles (corresponding to a count rate of the order
1000~s$^{-1}$) would enter into the main spectrometer. This
minimizes the chances of causing background by ionization of
residual gas and build up a trapped plasma in the spectrometer.
A filter width of $\Delta E \approx 50$~eV would be sufficient for
the pre-spectrometer. The flux tube in the analyzing plane would
then have a diameter of 1~m, corresponding to a field of 25~mT.

In a second application, the pre-spectrometer will act as a
 fast switch for running the main spectrometer in the non-integrating
 MAC-E-TOF mode (see section \ref{sec_macetof}).

 The pre-spectrometer of KATRIN will have a diameter of 1.2~m and
 a length of 3.5~m, so that its dimensions are comparable to the
 existing MAC-E Filters at Mainz and Troitsk. As the designs of the
 pre- and main spectrometer will be similar, the former
 will act as a test facility for the larger main spectrometer.
 Especially important will be the following tests of\,: a) the technique
 to achieve an XUHV of below $10^{-11}$\,mbar, b) the concept of
 using the vacuum vessel itself as main electrode, and c) the electrodynamic
 concepts to reduce background.

\begin{figure}
\epsfxsize=14cm
\centerline{\epsffile{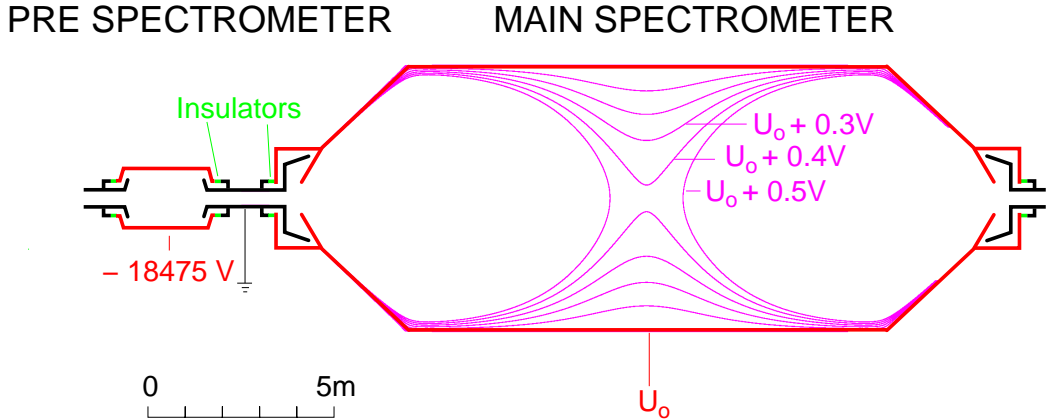}}
\caption{Schematic view of the pre- and main spectrometer. Shown are the
  electrostatic system and the vacuum vessels
  (black for ground potential, red
  for retarding potential and green for insulators),
  but not the magnets, since the diameter of the tubes
  at both ends of the spectrometer had to be enlarged
  in the drawing to become visible.
  From the equipotential lines  with 0.1~V
  difference (magenta) in the main spectrometer the electrical potential
  drop in the
  analyzing plane can be estimated to be 0.45~V.}
\label{fig_spectrometers}
\end{figure}

 \subsection{Main electrostatic spectrometer}

 A key component of the new experiment will be the large
 electrostatic spectrometer with a diameter of 7\,m and an overall
 length of about 20\,m. This high resolution MAC-E-Filter will allow
 to scan the tritium \bdec\ endpoint with increased
 luminosity at a resolution of 1\,eV, which is a factor of 4
 better than present MAC-E Filters.

\begin{figure}
\epsfclipon \epsfxsize=11cm
\centerline{\epsffile{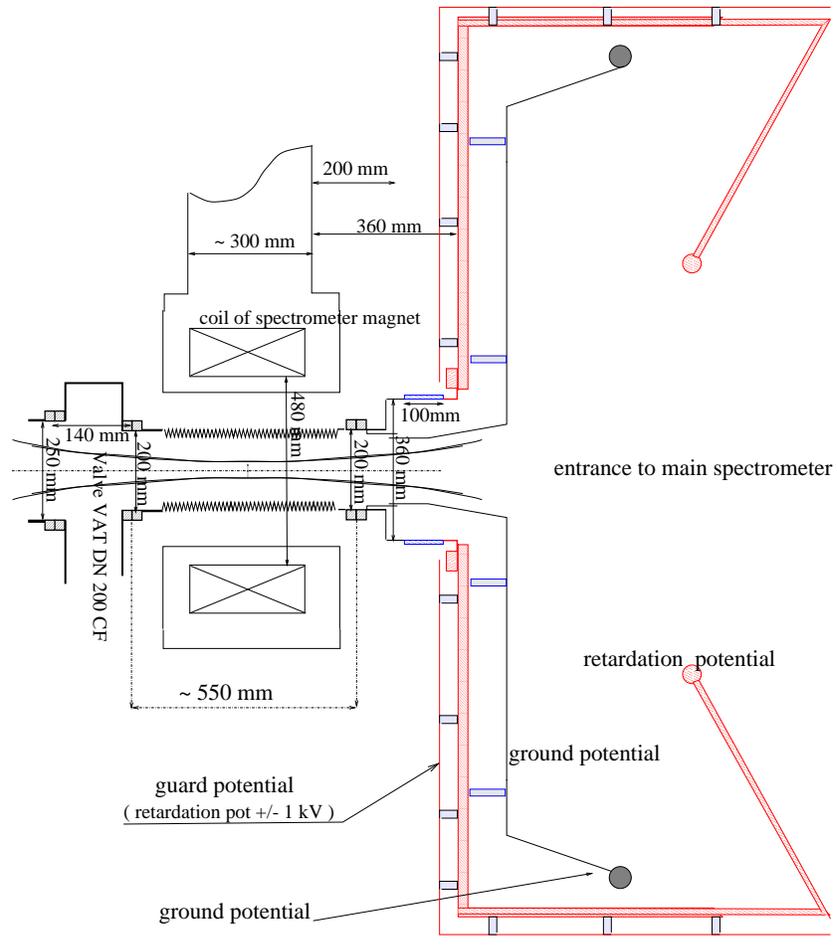}} \caption{Preliminary
design study of the end cap region of the main
 spectrometer.}
\label{fig_one_electrode}
\end{figure}

 The current default design is based on the concept that the
 vacuum tank itself serves as the main electrode of the spectrometer
 (see fig. \ref{fig_spectrometers}). This main electrode is connected
 on both sides by insulators to ground electrodes,
 which act at the same time as end-caps of the spectrometer vessel.
 To keep the inner HV system stable and safe, an
 outer HV shield at a guard potential close to
 the retarding potential will be
 installed. Figure \ref{fig_one_electrode} shows details
 of a preliminary design. The advantages of this concept
 is the minimization of degassing surfaces in the vacuum
 vessel and the much simplified construction work inside
 the vessel (as compared to the efforts for mounting and insulating
a huge electrode system within the vacuum vessel). In
 addition, one gains about 10~\% of cross section in the analyzing plane.

 The actual design of the vacuum vessel as well as the shape and mounting
 of the ground
 electrodes  is currently being optimized. The present version exhibits
 a tolerable potential change of about 0.4~V across the analyzing
 plane. The magnet design provides a homogeneity of the magnetic field in
 the analyzing plane of better than 10~\% . Both kind of inhomogeneities
 will be compensated by the good spatial resolution of the detector.
 The electrostatic
 optimization as well as the corresponding investigation and
 optimization of the magnetic fields and the simulation of the electron
 tracks are being pursued further by helps of dedicated 3 dimensional
 simulation programs.

 The combination of large tank dimensions together with the stringent XUHV
 requirements represents a technological challenge, as XUHV vessels of this size
 have not been manufactured previously. Therefore, the principal feasibility of
 the main spectrometer had to be investigated. A study in
 collaboration with several industrial
 partners has demonstrated the feasibility of the construction of a
 spectrometer of this size and has also yielded a concise
 construction schedule.
 During the construction phase special emphasis has to be put on
 surface preparation techniques. Especially important will be
 electro-polishing of inner tank surfaces to reduce field
 emission, and the baking of the vessel at temperatures  up to
 350$^{\circ}$\,C to reduce the outgassing rate.
 The latter task will require the installation of a
 heating system for the spectrometer with 100\,kW power.
 The heat will be transmitted by circulating oil through a tube system
 around the tank. This solution allows, moreover, to cool down the tank
 thereafter to a temperature of -20~$^\circ$C, where outgassing rates
 are expected to be at least one order of magnitude lower as compared to
 room temperature.
In combination, these techniques will allow to meet the aspired
degassing rate from the bulk
 material (UHV stainless steel) of $< 10^{-13}$\,mbar\,l\,s$^{-1}$\,cm$^{-2}$.
 The pumping system, which is specified according to this
 degassing rate, comprises a system of volume getter pumps
 (2.5~km of SAES getter strips)
 mounted inside the spectrometer vessel. In addition
 turbo molecular and ion getter pumps will be installed for pumping
 non-getterable residual gases ({\it e.g.} He). Altogether a
maximum pumping speed of $s = 5 \cdot 10^5$~l/s will be achieved for getterable
gases. This number bears an extra safety margin of
 an order of magnitude for reaching a final pressure  of
 $10^{-11}$\,mbar\footnote{An alternative concept would be to
 evaporate a titanium
 (or equivalent) getter inside the tank. This could provide not only a
 huge pumping speed but also a homogeneous,
 well controlled surface potential.}.
 The early tests with the pre-spectrometer vacuum system based on the same
 design will allow to optimize the final design.

\subsection{Non-integrating MAC-E-TOF Mode}
\label{sec_macetof}
In the recently developed MAC-E-TOF mode \cite{macetof} one adds on top of the
high pass filter of particle energy an equally sharp low pass filter by
additional time-of-flight (TOF) measurement through the spectrometer.
Both filters together form a narrow band pass.
Sufficiently precise TOF measurement is possible even for particles as light
as electrons since they are slowed down during their passage through the
analyzing potential.

The advantage of the
MAC-E-TOF mode is that it provides a non-integrating
spectrometer at same energy resolution and similar luminosity as the
MAC-E-Filter mode. Ideal applications for such a mode are the investigation
of systematic uncertainties like the precise determination of the inelastic
scattering cross section in the source. Also the shape of the tritium
\bspec\ near its endpoint can be investigated with high differential
resolution in order to search for new physics (apart from neutrino mass) like
tachyonic neutrinos \cite{ciborowski}
or small right-handed contributions to the electroweak interactions
\cite{stephenson}, etc.

In tritium $\beta$ decay the electron start time can only be
determined by chopping the flux. Pulsing the retarding potential
of the pre-spectrometer to a voltage above \ezero\ at a high
frequency of about 100~kHz will allow $\beta$ particles to enter
the main spectrometer only within short time windows. The HF
pulsing thus enables their time-of-flight analysis. The width of
the time windows will reduce somewhat the energy resolution and
luminosity. Simulations and experiments at the Mainz setup have
shown \cite{macetof} that in the MAC-E-TOF mode a triangle-like
energy resolution function of the same half width as the full
width of the transmission function of the integrating MAC-E-Filter
can be achieved at a loss of about a factor of 4 in the count
rate.

\subsection{Detector concept}

All $\beta$ particles passing the retarding potential of the
MAC-E-Filter will be guided by a magnetic transport system to the
detector. The detector requirements are the following:
\begin{itemize}
\item high efficiency for \el\ detection and simultaneously low $\gamma$
      background,
\item energy resolution of $\Delta E<600$\,eV
      for 18.6\,keV electrons to suppress
      background events at different energies,
\item operation at high magnetic fields,
\item position resolution to\\
      (i) map the source profile,\\
      (ii) localize
      the particle track within the spectrometer (for compensation of
      inhomogeneities of electric potential and magnetic field
      in the analyzing plane),\\
      (iii) suppress background originating
      outside
      the interesting magnetic flux ({\it e.g.} coming from
      the electrodes of the spectrometer),
\item for a measurement in a MAC-E-TOF mode, a reasonable time resolution
      ($\sigma_t<100$\,ns),
\item for test and calibration measurements ready to take high count rates
      (up to total rate of order 1~MHz)
\end{itemize}

The present concept of the detector is based on a large array of
silicon drift detectors. The array has to be sensitive over the
whole magnetic flux tube, corresponding to a diameter of 11 cm.
Silicon drift diodes are significantly advanced in energy and
spatial resolution over the rather simple detectors used in the
Mainz and Troitsk experiments. The silicon drift detectors will
have a very thin dead layer of only 50~nm in order to reduce
energy loss and straggling therein and thus to improve the energy
resolution. A thin sensitive layer of about 300~$\mu$m   will help
to reduce the $\gamma$ sensitivity. The small-sized readout
electrode -- the advantage of silicon drift diodes -- reduces
electronic noise. An energy resolution of $\Delta E = 600$~eV
(FWHM) for 18.6 keV electrons should be achieved.

The typical pixel size will be $3\times 3$\,mm$^2$ leading to
about 1000 read-out channels allowing a detailed source mapping
and reasonably low-sized electronic read-out. Furthermore, as
silicon drift diodes are semiconductor based detectors, they can
be operated in high magnetic fields of several T. Consequently,
the detector will be situated within a special solenoid generating
a magnetic field $B_D$ perpendicular to the detector surface. The
maximum angle $\theta_{\rm D,max}$ of incoming electrons at the
detector can be limited by the choice of field ratio (see eq.
\ref{eq_thetamax}) to
\begin{displaymath}
  \theta_{\rm D,max} \approx {\rm arc} \sin \left(\sqrt{\frac{\bd}{\bmax}}\right)
\end{displaymath}
Smaller angles $\theta$ reduce backscattering from the detector
surface and potential loss of electrons\footnote{Although most of
the backscattered electrons will be magnetically or
electrostatically reflected back onto the detector, the two
additional passages through the dead layer will shift a part of
the electron signal out of the predefined energy window.}.

\begin{figure}
\epsfxsize=13cm
\centerline{\epsffile{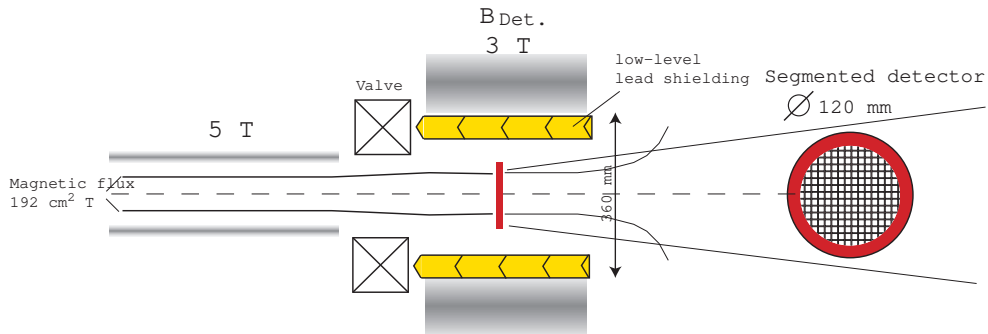}}
\caption{Schematic view of the detector area with magnet configuration and
  position of the electron detector.}
\label{det_schema}
\end{figure}

The detector will be surrounded by low-level passive shielding and
an active veto counter to reduce background.
Figure~\ref{det_schema} shows a view of the schematic setup of the
solenoids as well as the position of the detector surface.

For a later stage of the experiment a segmented bolometer is
considered as detector. This type of detector has the advantage of
a superior energy resolution (as compared to Si-detectors). For
cryogenic detectors, resolution values of \eg\ $\sigma_E\approx
5$\,eV at 60\,mK for short term measurements of X-rays of 6\,keV
have been reported \cite{imec}. In high magnetic fields the
detector performance was even improved. A better energy resolution
of the detector will play an important role, if the relevant
sources of background -in particular those from the spectrometer
itself (see the following section \ref{sec_background})- have an
energy distribution, which is sufficiently wider as (or separated
from) the energy interval close to the endpoint. In this context
it is important to note that recent data from the Mainz experiment
show that -within the statistical precision of about 25~eV- the
mean of the background line originating from the spectrometer
coincides with the filter potential. However, the limited energy
resolution of the detector at Mainz of $\Delta E = 1.4$~keV (FWHM)
does not allow the detection of a possible fine structure of the
energy distribution of the background events.

A cryogenic bolometer would need a proper 4$\pi$ heat shield at
LHe temperature. This would require another cold chicane for the
electron transport to the detector to prevent a direct line of
sight into the warm spectrometer. A significant amount of R\&D
work will be necessary to reach the energy resolution aimed at for
a whole array of medium size detectors.

More detailed studies are under way to define a final detector
design and setup as well as more quantified detector requirements.
However, arrays of Si drift diodes and bolometers represent two
promising concepts for the detection of low energy electrons in
KATRIN.

\subsection{Background}
\label{sec_background}

The experiments at Mainz and Troitsk observe nearly the same count
rate of background events of about 10~mHz from their
spectrometers, although the volume of the Troitsk spectrometer is
2 times larger and the residual gas pressure is 10 times higher.
This is not a contradiction to the principle of a strong
dependence of the background rate on the residual gas pressure in
the spectrometer, seen both at Mainz and Troitsk. It shows,
however, that the background rate also depends on other critical
parameters, such as {\it e.g.} the shape and strength of the
electric and magnetic fields. This fact and the detailed
background investigations at Mainz and Troitsk under different
conditions make us confident, that we will be able to reach about
the same level of background rate of 10~mHz also with the large
KATRIN spectrometer.

To allow a more detailed discussion, the different sources of
background are described in the following:
\begin{itemize}
  \item Environmental radioactivity and cosmic rays around the detector\\
        This background source
        can be reduced by using selected materials, by active and passive
        shielding and by a detector with good intrinsic energy resolution.
        At Mainz this background contributes with a rate of about 1~mHz.
        The same level can be obtained with the KATRIN detector.
        Although it will be 20 times larger in area, the
        background rate can be reduced by the same factor
        by the at least 2 times better energy resolution, an about
        2 times thinner active layer and a 5 times better shielding.
  \item Tritium decays in the main spectrometer\\
        In about 15~\% of all \bdec s of molecular tritium the remaining
        $^3$HeT$^+$ daughter ion gets further ionized and a low energy shake off
        electron accompanies the \belec . If the $\beta$ decay takes place in
        the detector-facing half of the spectrometer, the electron will be
        accelerated and collimated onto the detector. There
        it will be detected with an energy close to that of the retarding
        energy $qU_0$, which fills most of the spectrometer volume.
        Therefore it can not be distinguished from the signal electrons with
        present detectors. The only way to avoid
        this kind of background is a very stringent limit on
        the tritium partial pressure inside the main spectrometer.
        A \ttwo\ partial pressure of
        $p(\ttwo ) \simeq 10^{-20}$~mbar
        causes a background rate of 1~mHz, which is the maximum
        tolerable one from this source.
        The differential pumping and cryotrapping
        sections as well as the large pumping speed in pre and main
        spectrometer will guarantee this low \ttwo\ partial pressure.

        The $\beta$ particles from \ttwo\ decay within the spectrometer
        volume can contribute even stronger to this indistinguishable
        background by secondary reactions. Born at high starting angle
        in the low field \ba\,, they will be trapped in this magnetic bottle
        and cause ionizations of the residual gas \cite{Lobashev99}.
        However, this background can be suppressed by obeying XUHV conditions
        such that the chance of ionization is low within the storage time.
        The latter is limited by synchrotron radiation \cite{pic92a} and can
        be further shortened by breaking the cylindrical symmetry of the
        fields.
  \item Cosmic rays\\
        Cosmic rays can create secondary charged particles like
        $\delta$-electrons, which then emerge
        from the electrodes and penetrate into the spectrometer, where they
        are guided by the magnetic
        field. Low energy charged particles will stay always close to
        the outer walls of the spectrometer and can not reach the central magnetic flux,
        which is collimated onto the detector. These secondary charged
        particles therefore cannot contribute directly to the background.
        But they can create ions with much larger cyclotron orbits crossing
        the magnetic flux tube. Accelerated further by the electric
        field, these ions then have the chance of producing slow electrons by various
        tertiary reactions with the residual gas like ionization, charge
        exchange to negative ions with subsequent electron detachment etc.
        H$^-$ ions have eventually been
        observed at the detector. A chance to undergo
        such tertiary reactions under UHV conditions exists only for
        trapped particles, however.\\
        Measurements
        with the Troitsk and Mainz setup using an external $\gamma$-ray
        source to create $\delta$-electrons
        resulted in a suppression factor  of background electrons
        with respect to the primary produced $\delta$-electrons of
        about $10^{-6}$.
        The much larger dimensions and the much better vacuum
        of the KATRIN main spectrometer will improve this
        suppression factor and compensate the factor from
        the increased inner surface. Test experiments with the
        pre-spectrometer in 2002 will check whether cosmic rays
        are indeed of minor concern for the background.
  \item Trapped particles\\
        Measurements with the existing MAC-E-Filters have shown that
        the background rate can change abruptly, following time constants
        of minutes to hours and showing hysteresis like effects. This behavior
        must be due to trapped particles inside the spectrometer and
        chain reactions may sustain a plasma (like in a penning gauge)
        and release electrons to the detector. In principle, MAC-E-Filters
        can have phase space regions in which
        particles of either charge can be trapped, particularly in case of an
        awkward field design.
        The hypothesis that traps contribute to the background is further
        supported by the fact that
        at Mainz ``heating'' of the trapped particles
        by a high frequency electric AC field decreases and stabilizes the
        background count rate.\

        Although the KATRIN spectrometer has a much larger volume
        than the present spectrometers, the increase of dimensions
        and the careful electrostatic and magnetic design
        will also allow to reduce the electric fields and
        field emission from surfaces.
        Furthermore the design will improve the magnetic
        shielding of the central magnetic flux from charged particles
        originating at the electrodes or walls.
        In addition, new concepts of active background suppression
        will be tested.
        To avoid the trapping of charged
        particles in regions corresponding to
        magnetic bottles and the resulting chain of
        ionization processes, the axial field symmetry may be
        broken by bending of the magnetic flux tube. Alternatively, electric
       dipole fields may be applied. The resulting transverse $\vec{R} \times
\vec{B}$ drift (with $\vec{R}$ defining the radial vector) in a
non-straight geometry or $\vec{E} \times \vec{B}$ drift in the
case of an electric field perpendicular to the magnetic field
lines should remove stored particles from the sensitive flux tube
on time scales of less than 1~s.
 Such investigations are foreseen at the present
        spectrometers as well as at the pre-spectrometer. In parallel, ray tracing
        of particles on the computer will minimize trapping possibilities
        by optimized field design and improve our understanding of these
        processes.
\end{itemize}

\subsection{Systematic uncertainties}
\label{sec_systematics}

For a high sensitivity tritium $\beta$ experiment like KATRIN, the
interesting region of the $\beta$ spectrum close to the endpoint
$E_0$  is very narrow. A narrow energy interval means that the
count rate statistics will be limited, i.e. that the statistical
error is rather large. On the other hand, a narrow energy interval
strongly reduces systematic errors. The systematic uncertainties
of the current tritium $\beta$ experiments mainly arise from
processes connected to atomic and molecular physics, such as
inelastic scattering of tritium \belec s in the tritium source.
This process as well as various other sources of systematic
uncertainties are discussed in more detail below.

\begin{itemize}
  \item Inelastic scattering\\
    The inelastic scattering of electrons in the tritium source
    is one of the dominant sources of the systematic background.
    The cross section of this process \cite{Aseev} has a rather
    high threshold of more than 12~eV (see fig.
    \ref{fig_response}). Thus, the last 12~eV of the \bspec , which
    carry the main information
    on the $\nu$ mass, are free of any inelastic scattering events. This holds also
    strictly for the measured spectrum, because the transmission
function of a MAC-E-Filter has no high energy tail (see fig.
\ref{fig_response}) due to energy conservation.

    Confining the measurements to the elastic fraction, much thicker
    tritium sources can be used.
    The present relative uncertainties of the total
    cross section are 2~\% for gaseous tritium and 5~\% for quench condensed
    tritium sources
    \cite{Aseev}. The shape of the energy loss function is well
    measured for gaseous tritium and reasonably well known for
    quench condensed tritium as illustrated in fig. \ref{fig_response}.
    Still the knowledge of
    both the total as well as the differential cross section
    can be improved further by dedicated measurements
    with quasi-monoenergetic electron sources
    (electron gun or K-32 conversion from \kr ) using the MAC-E-TOF mode of the
    new KATRIN spectrometer. Residual systematic uncertainties will be reduced
    accordingly.
  \item Column density and homogeneity of the tritium source\\
    The column density of the WGTS and the QCTS is measured and monitored
    in two ways\,:\\ (i) The tritium count rate determined with the detector
    gives online a spatially resolved tritium column density measurement
    in combination with online mass spectrometry.
    For this purpose a particular measurement point further below the tritium
    endpoint with enhanced count rate will be chosen.\\ (ii)
    Offline measurements with the electron gun will control the column density
    of the WGTS from the ratio of the inelastic to elastic
    fraction,
    whereas that of the QCTS is determined by ellipsometry.
    A precision of 1~\%  in units of the mean free column density
    $(\rho d)_{\rm free}$
    can be achieved safely.
  \item Backscattering\\
    The coefficient describing the fraction of electrons backscattered
    from the graphite substrate of the Mainz experiment is
    $3 \cdot 10^{-5}$/eV and even smaller for the Troitsk setup. Therefore,
    backscattering does not play any significant role for the
    narrow
    energy interval below the $\beta$ endpoint
    considered for the KATRIN experiment.
  \item Final states\\
    The first electronic excited state of the $^3$HeT$^+$ daughter molecule has
    an excitation energy of 27~eV \cite{saenz1}.
    Therefore excited states do not play
    any role for the  energy interval considered for the KATRIN experiment,
    only the decay to the ground state
    of the ($^3$HeT)$^+$ daughter molecule, which is populated with 57~\%
    probability,  has to be taken into account.
    Due to the nuclear recoil, however,
    a large number of rotational-vibrational
    states with a mean excitation energy of 1.7~eV and a width of 0.4~eV
    is populated. This distribution follows the Franck-Condon principle;
    its precision depends on the
    knowledge of the ground state wave function, which is extremely good
    \cite{saenz2}.
    Therefore, no significant uncertainty arises from the
    rotational-vibrational
    excitation of the final ground state. Also a contamination of the \ttwo\
    molecules by HT does not matter in first order: The shift of the mean
    rotational-vibrational excitation of HT with respect to \ttwo\ is
    compensated by a corresponding change of the nuclear recoil energy of HT
    with respect to the 1.5 times heavier \ttwo\ molecule \cite{saenz1}.
    However, this distribution ultimately limits the resolution which
    can be obtained in \ttwo\ \bdec\,.
  \item Transmission function\\
    Since the transmission function (eq. \ref{eq_mace}) depends only
    on the
    relative field values and potentials, it is insensitive to
    mechanical adjustment. The inhomogeneity of the magnetic field and
    the electric potential can be calculated precisely. In addition,
    the shape of the transmission function will be checked by
    a point-like test source of K-32 conversion electrons from \kr\
    or by an electron gun, which is moved across the magnetic flux
    tube at the position of the QCTS.
    \\
    Fluctuations of the absolute position of the transmission function are
    more critical. A simple relation connects an additional unknown Gaussian
    broadening of
    width $\sigma_g$ ({\it e.g.-} caused by fluctuations of the absolute value of the
    retardation potential $qU_0$) to a shift of \mtwo\ \cite{robertson}:
    \begin{equation}
      \dmtwo = - 2 \cdot \sigma_g^2
    \end{equation}
    Therefore, the noise and the stability of the high voltage has to be
    below 70~mV to limit its contribution to \dmtwo\ to a maximum
    value of 0.01~eV$^2$.
  \item Trapped electrons in the WGTS\\
    Each \belec\ which is trapped in a local magnetic minimum due to eq.
    (\ref{eq_thetamax}) will suffer inelastic scattering events.
    Rarely these processes are accompanied by a momentum transfer large enough to
    scatter the electrons into the cone of polar angles small enough to
    leave the trap. Before being freed finally, they hence accumulate an energy
    loss being larger, most probably, than the energy region of interest
    below the endpoint.
    Furthermore the magnets will be designed to avoid large trapping volumes
    in the WGTS.  Magnetic traps cannot be avoided in the region of differential
    pumping ports (see fig. \ref{difpump}). In the bent, however, the electrons
    will be driven out of the trap by synchrotron radiation.
    Moreover,
    the \ttwo\ density and hence the decay rate is dropping very fast along
    the differential pumping chain.
  \item Solid state effects for the QCTS\\
     Several additional systematic uncertainties are connected with the
     QCTS:
     \begin{itemize}
       \item Neighbor excitation\\
         The sudden change of nuclear charge in $\beta$ decay can excite
         even neighboring \ttwo\ molecules. According to ref.  \cite{kolos}
         the probability is 5.9~\% in a closely packed single crystal and
         the mean excitation energy is given as 14.6~eV, based on the spectrum
         of free hydrogen molecules. For the analysis of the Mainz data the
         former number has been lowered to 4.6~\% and the latter raised to
         16.1~eV \cite{Weinheimer99}. The changes account for the reduced
         density of the quench condensed film \cite{Weinheimer99}
         and for the observed shift of the energy loss spectrum of 18~keV
         electrons passing gaseous and solid hydrogen, respectively
         \cite{Aseev}. This shift is also corroborated by quantum-chemical
         calculations \cite{Saenz4}.
         To be conservative the changes are taken fully into account as
         systematic uncertainties \cite{Weinheimer99}. To improve on
         this situation
         a quantum chemical calculation of the neighbor excitation for the
         quench condensed \ttwo\ case would be useful. For the KATRIN data
         this effect will play a marginal role, since inelastic events
         will do not contribute to the signal in a significant
         way.
       \item Self-charging\\
         Due to the continuous radiation of $\beta$ particles
         a quench condensed \ttwo\ film
         -- being an excellent insulator -- charges up.
         This effect has been studied in detail and explained by
         a satisfactory model \cite{Erice,bornschein2001}. In the temporal
         equilibrium self-charging generates a linear drop of the electric
         potential across the \ttwo\ film
         of about 20~mV per monolayer. For the Mainz analysis \cite{Bonn2000}
         20~\% of the total self-charging effect was taken into account
         as systematic uncertainty. Refined measurements could probably
         reduce this uncertainty by another factor of 2. More serious than
         systematic uncertainties is the 2~eV broadening of the spectrum due to
         self-charging of the QCTS since it reduces the sensitivity on \mnu\
         (see below).
       \item Longterm behavior\\
         A tritium loss of about 0.16 monolayer per day
         was observed at the Mainz QCTS. The effect is due to
         sputtering of \ttwo\ molecules
         by nuclei recoiling from \bdec s. It cannot be avoided
         but has been monitored precisely by measuring the
         longterm decrease of the
         count rate. In parallel a condensation of \htwo\ from the residual
         gas was observed by ellipsometry at the end of the run.
         This effect will be avoided
         in the KATRIN experiment by providing
         more effective cryogenic vacuum conditions at the QCTS.
     \end{itemize}
\end{itemize}
In summary, the main systematic uncertainties arise from the
inelastic scattering and the degree of stability of the retarding
voltage. For the case of the QCTS source the uncertainties of the
solid states effects have to be added. At KATRIN the systematic
uncertainties will be substantially reduced by
\begin{itemize}
  \item the much smaller energy interval of interest
        below the endpoint \ezero ,
  \item additional control measurements at ultrahigh resolution in both the MAC-E- and
        MAC-E-TOF mode,
  \item the better stability of the high voltage,
  \item the higher \ttwo\ concentration and
  \item the improved vacuum conditions
\end{itemize}
with respect to the present experiments. As will be shown in
section \ref{sec_sensitivity}, this improvement will allow KATRIN
to reach the goal of a sub-eV sensitivity on the neutrino mass.

\subsection{High voltage stability and energy calibration}
\label{sec_stability} As pointed out in section
\ref{sec_systematics}, a high voltage stability of better than
70~mV is needed. This requirement corresponds to a precision of a
few ppm, which is state of the art for a system consisting of a
high voltage (HV) power supply, a HV divider and a digital
voltmeter. The longterm stability of this system will be checked
repeatedly by measuring the K-32 conversion line of \kr , which
has an energy of about 17.825~keV and a width of 2.83~eV (FWHM)
\cite{pic92b}. Such a measurement can be performed by quench
condensing krypton onto the graphite substrate of the QCTS. For
the WGTS this would require a 100 times larger  amount of krypton
circulating in the closed pumping cycle. Also the temperature has
to be raised from 30~K to about 70~K to avoid condensation of
krypton. One can also consider to use a third independent
electrostatic spectrometer, like the Mainz MAC-E-Filter, for
monitoring of the high voltage by measuring the K-32 conversion
line continuously. Since the half life of \kr\ is 1.83 hours only,
one would install the mother $^{83}$Rb (half life: 86 days) as a
source, continuously producing  \kr\ in situ. First investigations
of this technique look promising \cite{kovalik} but chemical and
solid state effects on the conversion line position have to be
thoroughly examined. Since the HV dividers and digital voltmeters
are generally more stable than the HV power supplies themselves,
the quality of the $\beta$ spectrum accumulated during long
measuring times can be improved by the method \cite{dragoun95}.

Besides stability a precise, absolute calibration of the
spectrometer voltage would be very valuable as an important check,
since the endpoint energy \ezero , derived from the measured
\bspec , could be compared to the independently measured mass
difference $\Delta M(^3{\rm He},{\rm T})$, which is known from
measurements with absolute precision of 1.7~eV \cite{vandyke}.
This measurement from the early nineties could be improved by
about one order of magnitude using state of the art equipment
\cite{Quint}. Then the comparison at a 0.1~eV level of the
experimental endpoint energy \ezero\ with the external mass
difference  $\Delta M(^3{\rm He},{\rm T})$ will check for unknown
systematics of the KATRIN experiment. The option to use the
external mass difference $\Delta M(^3{\rm He},{\rm T})$ directly
as fixed input parameter in the analysis of the \bspec\ to improve
its sensitivity on \mnu\ will be discussed in section
\ref{sec_sensitivity}.

\subsection{Sensitivity on the electron neutrino mass}
\label{sec_sensitivity}

For the calculation of the expected sensitivity on the electron
neutrino mass we restrict the discussion to simulated data from
the WGTS. As shown in section \ref{sec_parameters}, the signal
rate $S$ of electrons near the endpoint \ezero\ (see eq.
(\ref{eq_signal})--(\ref{eq_s_with_lambdafree})) can be expressed
in terms of the energy resolution $\Delta E/E$, the column density
$\rho d$ and the maximum accepted starting angle \thmax . Apart
from the count rate or signal strength near the endpoint,
sensitive parameters of the neutrino mass evaluation are the
background rate $b$, the actual interval below \ezero\ used for
measuring and fitting the spectrum, as well as systematic
uncertainties discussed in sec. \ref{sec_systematics}. As pointed
out there, the dominant systematic uncertainty using the WGTS is
inelastic scattering. In our discussion we consider this
uncertainty as the only one. We  used the present uncertainties,
characteristic for the Troitsk experiment \cite{Aseev}, although
an improvement by a factor of two at KATRIN can be expected.

\begin{figure}[tb]
\epsfxsize=10cm
\centerline{\epsffile{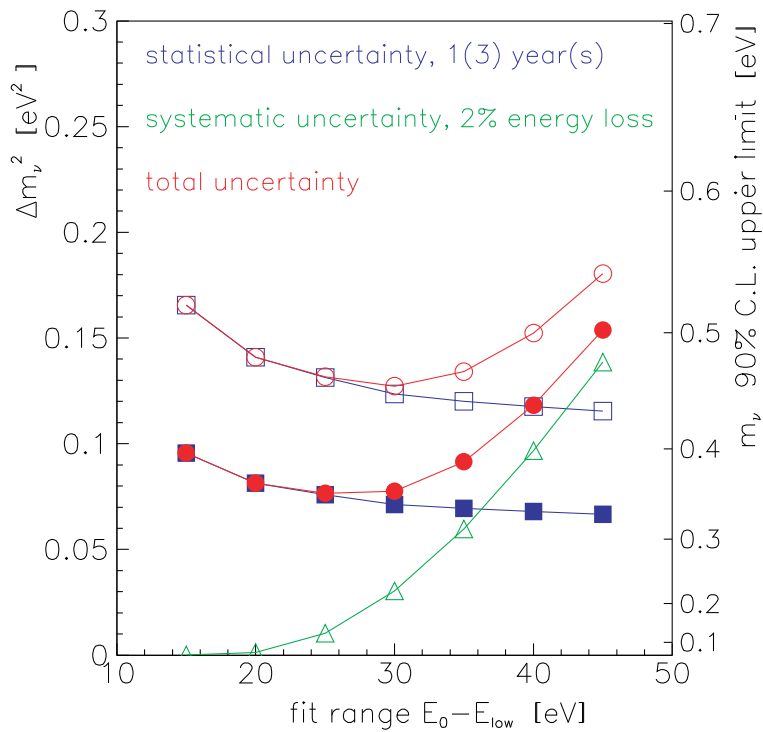}}
\caption{Estimates of the sensitivity on the electron neutrino mass for a
specific set of parameters (see text for details) shown as a function of the
fit range of the  electron spectrum below the endpoint \ezero .
The left axis denotes the $1 \sigma$ error of $m_{\nu}^2$,
the right axis denotes the $1.64\sigma$ or 90\% confidence upper limit on
the absolute electron neutrino mass under the assumption of zero mass.
Squares describe statistical uncertainties after 1 and 3 years
(open and filled squares,
respectively), triangles show the systematic uncertainty and circles represent
the total uncertainty.}
\label{fig_sensitivity}
\end{figure}

Fig.~\ref{fig_sensitivity} shows the results of simulations based
on a realistic parameter set: energy resolution $\Delta E=1$\,eV,
source surface area \as = 29\,cm$^2$, column density $\rho
d=5\cdot 10^{17}$/cm$^2$, maximum accepted angle $\thmax =
51^{\circ}$ for \belec s starting from the source, and  total
background rate of 11\,mHz. For small fit ranges near the
endpoint, the sensitivity is limited by statistics whereas for
larger intervals systematic uncertainties become dominant. After
one (three) year(s) of measuring time, the $1 \sigma$ error on the
observable \dmtwo\ is expected to be smaller than 0.14\,eV$^2$
(0.08\,eV$^2$), as can be seen from the left ordinate in
fig.~\ref{fig_sensitivity}. Having simulated the spectra with zero
neutrino mass, the one year measurement leads  to an upper limit
of the mass itself of 0.5\,eV at 90\% confidence. This upper limit
L(90\%) is connected to the error on \dmtwo\ via
$L(90\%)=\sqrt{1.64\cdot \Delta \mtwo}$. After three years of
measuring time, this limit will improve to
\begin{displaymath}
        \mnue < 0.35~ eV \qquad \mbox{at 90\,\% confidence},
\end{displaymath}
if no finite neutrino mass is observed. This sensitivity improves
the existing limits by almost one order of magnitude and also
demonstrates the discovery potential of KATRIN for an electron
neutrino mass in the sub-eV range.

The sensitivity estimates given above are based on the assumption
that KATRIN reaches the same background level as the present Mainz
and Troitsk experiments. As pointed out in section
\ref{sec_background} this assumption is reasonable. Significantly
larger or smaller rates of background would not change the
sensitivity of KATRIN to \mnue\ very much. If the background rate
would be 5 times higher, the limit on \mnue\ would be worse by a
factor 1.4, if the background would be 5 times lower, the limit on
\mnue\ would improve by a factor 1.2\,.

The sensitivity which can be obtained by measurements with the
QCTS, assuming the uncertainties of the Mainz experiment, will be
2 times worse than the WGTS sensitivity. The main disadvantage is
the self-charging, which will limit the signal rate $S$ as well as
the energy resolution $\Delta E$ as pointed out in section
\ref{sec_qcts}. If the self-charging effect cannot be overcome,
the QCTS will still provide checks with well understood and
complementary systematic uncertainties and as a backup source,
moreover.

In the standard fit procedure the four free parameters are
\mtwo\,, \ezero, amplitude $A$ and background rate $b$. As
mentioned in section \ref{sec_stability}, the endpoint energy
\ezero\ can be obtained from an independent measurement of the
$^3$He-T mass difference $\Delta M(^3{\rm He},{\rm T})$. Does it
make sense to put this number into the analysis as external fixed
input parameter for \ezero , in order to decrease the number of
highly correlated fit parameters by one? The correlation of \mnu\
to \ezero\ in the fit increases linearly with increasing the
distance to the endpoint\cite{otten_erice93}, which is the
neutrino energy. Therefore we have to restrict ourselves to
measure only a very small interval below \ezero\ of the last 5~eV.
Taking into account the width of the transmission function $\Delta
E$ and the average rotational vibrational excitation of the ground
state of 1.7~eV the effective endpoint is about 2.2~eV lower, thus
the interval quoted corresponds essentially to about the last 3~eV
of the spectrum. Fixing the endpoint energy the statistical
uncertainty of \mtwo\ for a 3 years measurement under standard
conditions (see above) becomes \dmtwo\ = 0.05~eV$^2$ . The main
systematic uncertainty will come from the external endpoint
energy, which results for this situation in $\dmtwo\ / \Delta
\ezero \approx 2$~eV. {\it E.g.} an uncertainty of $\Delta \ezero
= 20$~meV corresponds to an systematic uncertainty of
0.04~eV$^2$\,. Thus, if a 1~ppm precision in the $^3$He-T mass
difference $\Delta M(^3{\rm He},{\rm T})$ and the absolute
calibration of KATRIN could be achieved the sensitivity on \mnu\
could be improved further by using an external $\Delta M(^3{\rm
He},{\rm T})$ value in the analysis.

\section{Outlook and Conclusion}

{\large \bf Outlook}\vspace{0.4cm}

We have demonstrated the feasibility of a tritium \bdec\
experiment with sub-eV sensitivity. The realization of the
proposed KATRIN experiment will, nevertheless, be a technological
challenge, especially with regard to the extreme UHV requirements
of the large electrostatic spectrometer vessel.  In this context,
the existing infrastructure and technical expertise of
Forschungszentrum Karlsruhe will make it the favorable location
for the experiment. The overall costs of the experiment are
estimated to be about 17 million Euros (salaries not included),
with the main costs arising from the spectrometer vessel
(3.5\,M\,Euros estimated) and the solenoid system (6.7\,M\,Euros
estimated).

Following the first presentation of the KATRIN project at an
international workshop at Bad Liebenzell \cite{liebe}, the KATRIN
Collaboration was formally founded in June 2001. The future time
schedule of the KATRIN project calls for a full proposal to be
submitted in early 2002, followed by requests to the funding
agencies later that year. While systematic studies of background
processes have already been performed at Troitsk and Mainz,
further detailed studies as well as prototyping experiments and
further design optimizations will be carried out over the next
months. On condition that the funding requests will be approved,
the construction works for KATRIN could be finished by the end of
2005. The commissioning and first test measurements of KATRIN
could then start in 2006, with long term data taking starting
later that year. We aim for a strong collaboration to build and
perform the proposed experiment and would welcome new
participants.
\vspace{0.6cm}\\
{\large \bf Conclusion}\vspace{0.4cm}

In this paper we have discussed the physics and technique of a
next generation tritium \bdec\ experiment, which would have an
unprecedented sensitivity to the electron neutrino mass. The
experiment we propose has the potential to improve the
sensitivities of the present experiments in Troitsk and Mainz by
{\it one order of magnitude}. With an estimated sensitivity of  $m
(\nue)$ = 0.35 eV\,, KATRIN could investigate for the first time
the sub-eV neutrino mass range by a direct kinematic measurement.

Neutrino masses are of special interest for cosmology and
astrophysics, as the relic neutrinos left over from the Big Bang
could play a significant role as neutrino hot dark matter in the
evolution of large scale structures in the universe. If the
electron neutrino mass falls into the sensitivity range of KATRIN,
the role of relic neutrinos in structure formation could be fixed
(taking into account the recent results of neutrino oscillation
experiments). If, on the other hand, the $\nue$-mass will not be
in the range of the sensitivity, the constraint on the
contribution $\Omega_{\nu}$ of relic neutrinos to the total
matter-energy density of the universe could be improved by one
order of magnitude, thereby limiting the cosmological significance
of neutrinos. Therefore, one of the main motivations of KATRIN is
to measure or to constrain the parameter $\Omega_{\nu}$. In this
context, the results of KATRIN would be an important input
parameter for cosmological studies, especially for the analysis of
future high precision, satellite based cosmic microwave background
measurements (MAP and Planck).

Apart from astrophysics and cosmology, the absolute mass scale of neutrinos plays
a central role for particle physics. As the Standard Models of particle physics offers
no explanation for fermion masses and mixing, the determination of the $\nu$-mass
scale will be of fundamental importance for extended theories beyond the Standard
Model dealing with the mechanisms of mass generation. As $\nu$ masses are much
smaller than the masses of the other fermions, it will most probably be the neutrino mass
scale which
will set the scale for new physics.

The only method to investigate the absolute mass scale of
neutrinos in a {\it model independent} way is the high precision
spectroscopy of \bdec s. For this class of experiments, only
tritium \bdec\ experiments using electrostatic spectrometers will
allow to reach the sub-eV mass range in the nearer future. The
experiment we propose will push this technique to its
technological limits, especially with regard to the dimensions of
the electrostatic spectrometer and the source strength of the
gaseous molecular tritium source. Thus, according to our present
knowledge, KATRIN represents an 'ultimate' tritium \bdec\
experiment.

The proposed next-generation tritium \bdec\ experiment will be complementary
to the numerous future oscillation experiments using solar, atmospheric and accelerator
neutrinos. These experiments will determine with great precision the neutrino mixing
parameters as well as the mass splittings among the different neutrino mass eigenstates,
but will not yield information on the absolute values of neutrino masses.
KATRIN will also be complementary to future \nnbb\ experiments, which will provide important
information on Majorana neutrino masses. It is only the combination of the different results
from neutrino oscillation experiments, \nnbb\ experiments and tritium \bdec\
experiments which will allow us to get the 'full picture' of neutrinos.

\end{document}